\documentclass[journal,10pt,twocolumn,a4]{IEEEtran}
\usepackage{cite}
\usepackage[cmex10]{amsmath}
\usepackage{amsbsy}
\usepackage{algorithm,algorithmic}
\usepackage{tikz}
\usetikzlibrary{shapes, shadows, arrows}
\usepackage[hyphens]{url}

\usepackage{graphicx}
\usepackage{float}
\usepackage{stfloats}
\usepackage{wrapfig}

\usepackage{epstopdf}

\usepackage[latin1]{inputenc}

\usepackage{amsthm,dcolumn,graphicx,multicol,fancyhdr}
\usepackage{latexsym}
\usepackage{graphicx}
\usepackage{ifthen}
\usepackage{rotating,amsfonts,color,amssymb,amsmath,amscd,array,enumerate,afterpage}

\usepackage{hyperref}
\usepackage{multirow}

\usepackage{enumitem}
\setlist{parsep=3pt,listparindent=\parindent}

\usepackage[caption=false,font=footnotesize]{subfig}

\theoremstyle{definition}

\def\subfigure{\subfloat}

\newcommand{\babc}{\renewcommand{\labelenumi}{(\alph{enumi})}\begin{enumerate}}
\newcommand{\eabc}{\end{enumerate}}
\newcommand{\biii}{\renewcommand{\labelenumi}{(\roman{enumi})}\begin{enumerate}}
\newcommand{\eiii}{\end{enumerate}}

\newcommand{\beqn}{\begin{eqnarray*}}
\newcommand{\beq}{\begin{eqnarray}}
\newcommand{\eeqn}{\end{eqnarray*}}
\newcommand{\eeq}{\end{eqnarray}}

\DeclareMathOperator* {\argmin}{arg\,min}

\DeclareMathOperator* {\sign}{sign}

\newcommand{\ckboldon}[1]{#1}

\newcommand{\ckbold}[1]{%
 \ifthenelse{\isundefined{\ckboldon}}{#1}{ \textbf{#1} }
}

\newcommand{\tr}{\mbox{tr}\,}

\begin{document}

\setlength{\abovedisplayskip}{7pt}
\setlength{\belowdisplayskip}{7pt}
\raggedbottom

\title{Separating Stimulus-Induced and Background Components of Dynamic Functional Connectivity in Naturalistic fMRI}

\author{Chee-Ming Ting, \IEEEmembership{Senior Member, IEEE}, Jeremy I. Skipper, Steven L. Small and Hernando Ombao\vspace{-0.33in}
\thanks{C.-M. Ting is with the School of Information Technology, Monash University Malaysia, 47500 Bandar Sunway, Selangor, Malaysia, and also the Biostatistics Group, King Abdullah University of Science and Technology, Thuwal 23955, Saudi Arabia (e-mail: ting.cheeming@monash.edu).}
\thanks{J.I. Skipper is with the Department of Experimental Psychology, University College London, UK (e-mail: jeremy.skipper@ucl.ac.uk).}
\thanks{S.L. Small is with the School of Behavioral and Brain Sciences, University of Texas at Dallas, USA (e-mail: small@utdallas.edu).}\thanks{H. Ombao, is with the Biostatistics Group, King Abdullah University of Science and Technology, Thuwal 23955, Saudi Arabia (e-mail: hernando.ombao@kaust.edu.sa).}
}

\markboth{}%
{Shell \MakeLowercase{\textit{et al.}}: Bare Demo of IEEEtran.cls for Journals}

\maketitle

\begin{abstract}
We consider the challenges in extracting stimulus-related neural dynamics from other intrinsic processes and noise in naturalistic functional magnetic resonance imaging (fMRI). Most studies rely on inter-subject correlations (ISC) of low-level regional activity and neglect varying responses in individuals. We propose a novel, data-driven approach based on low-rank plus sparse (L+S) decomposition to isolate stimulus-driven dynamic changes in brain functional connectivity (FC) from the background noise, by exploiting shared network structure among subjects receiving the same naturalistic stimuli. The time-resolved multi-subject FC matrices are modeled as a sum of a low-rank component of correlated FC patterns across subjects, and a sparse component of subject-specific, idiosyncratic background activities.
To recover the shared low-rank subspace, we introduce a fused version of principal component pursuit (PCP) by adding a fusion-type penalty on the differences between the rows of the low-rank matrix. The method improves the detection of stimulus-induced group-level homogeneity in the FC profile while capturing inter-subject variability. We develop an efficient algorithm via a linearized alternating direction method of multipliers to solve the fused-PCP. Simulations show accurate recovery by the fused-PCP even when a large fraction of FC edges are severely corrupted. When applied to natural fMRI data, our method reveals FC changes that were time-locked to auditory processing during movie watching, with dynamic engagement of sensorimotor systems for speech-in-noise. It also provides a better mapping to auditory content in the movie than ISC.
\end{abstract}


\vspace{-0.15in}

\section{Introduction}

\IEEEPARstart{N}{aturalistic} functional magnetic resonance imaging (fMRI) is an emerging approach in cognitive neuroscience, employing naturalistic stimuli (e.g., movies, spoken narratives, music, etc.) to provide an ecologically-valid paradigm that mimics real life scenarios \cite{Sonkusare2019}. Complementary to traditional task-based paradigms with strictly controlled artificial stimuli, naturalistic paradigms offer a better understanding of neural processing and network interactions in realistic contexts, where stimuli typically engage the brain in continuous integration of dynamic streams of multimodal (e.g., audiovisual) information. Resting-state fMRI, although entirely unconstrained, is vulnerable to confounds and difficult to link with ongoing cognitive states. Natural stimuli like movies have shown higher test-retest reliability than resting and task fMRI \cite{Wang2017}. In this paper, we focus on a major challenge in naturalistic fMRI data analysis -- to extract stimulus-related neural dynamics from other intrinsic and noise contributions.

Early studies use traditional methods for task fMRI such as the general linear model (GLM) to detect brain activation driven by natural stimuli, using annotations of specific features or events (e.g., faces, scenes or speech in a film \cite{Bartels2004}) as regressors. However, GLM-based analysis requires explicit \textit{a priori} models of relevant features of the stimuli and associated neural responses as predictors, where minor model misspecification can result in low detection power. It is therefore effective only for simple parametric designs, but extremely rigid for capturing the richer dynamical information present in natural stimuli. Inter-subject correlation (ISC) analyses provide a powerful, data-driven alternative for handling naturalistic paradigms without a pre-defined response model, by leveraging the reliable, shared neural responses across different subjects when exposed to the same continuous naturalistic stimulation \cite{Hasson2004,Hasson2010}. By correlating fMRI time courses from the same regions across subjects, ISC can detect inter-subject synchronization of brain activity in specific areas activated by the stimuli, e.g., prefrontal cortex during movie viewing \cite{Jaaskelainen2008}. It was recently extended to inter-subject functional correlation (ISFC) to characterize stimulus-related functional networks by computing the correlations between all pairs of regions across subjects \cite{Simony2016}. Use of ISFC has revealed, for example, engagement of default mode networks during narrative comprehension \cite{Simony2016}. The ISFC increases specificity to stimulus-locked inter-regional correlations, by filtering out intrinsic, task-unrelated neural dynamics as well as non-neuronal artifacts (e.g., respiratory rate; head motion) that are theoretically uncorrelated across subjects. Time-resolved ISFC computed using a sliding-window technique has been used to track movie-induced dynamic changes in functional connectivity \cite{Bolton2018}. A recent study \cite{Di2020} explored ISC of time-courses of dynamic connectivity rather than regional activity. For a review of the ISC family of approaches see \cite{Nastase2019}. However, one limitation of IS(F)C is that it has no inherent way to detect individual differences in neural activity, due to the averaging-out of all uncorrelated variations across subjects. Moreover, the ISFC analysis inherently relies on the Pearson correlations between activity time courses of subjects, and thus how it can be generalized to other measures of brain connectivity, e.g., for directed functional networks is still unclear.

We propose a new approach based on low-rank plus sparse (L+S) decomposition to separate stimulus-induced and background components of dynamic functional connectivity (dFC) in naturalistic fMRI. Our method is inspired by the successful applications of L+S decomposition in video surveillance to separate a slowly-changing background (modeled by a low-rank subspace) from moving foreground objects (sparse outliers)
from a video sequence \cite{Bouwmans2014,Liu2015}. We formulate a time-varying L+S model for dFC, where the time-resolved, multi-subject functional connectivity (FC) matrices are represented as the sum of a low-rank component and a sparse component. The low-rank component corresponds to the similar or correlated connectivity patterns across subjects elicited by the same stimuli, which can be well-approximated by a common low-dimensional subspace. The sparse component captures the corruptions on stimulus-related FC edges, arising from background activity (i.e., intrinsic processes and artifacts) that are idiosyncratic and occur occasionally in certain subjects. We apply the robust principal component analysis (RPCA) \cite{Wright2009,Candes2011} at each time point to recover a sequence of time-resolved low-rank FC matrices from the sparse background noise, by solving a convex optimization that minimizes a combination of nuclear norm and $\ell_1$ norm (called principal component pursuit (PCP)). Under mild assumptions (i.e., the sparse matrix is sufficiently sparse), the RPCA-PCP can exactly recover the shared low-rank FC structure across subjects, even though an unknown fraction of the FC edges are corrupted by artifacts of arbitrarily large magnitude, as shown in \cite{Candes2011} for the general case. A related work \cite{Aggarwal2018} imposed L+S constraints on FC matrices of individual subjects and focused on modeling raw fMRI signals, another \cite{Jiang2020} used a low-rank approximation of stacked FC networks across subjects. However, both studies are limited to analyses of static FC and resting-state data.

We further introduce a novel L+S decomposition method called the fused PCP, by adding a new fusion-type penalty in the PCP to shrink the successive differences between the columns of the low-rank matrix towards zero. The fused PCP encourages homogeneity in connectivity patterns across subjects, by smoothing out the remaining stimulus-unrelated individual differences in the common connectivity structure. Moreover, it can provide a better recovery of the low-rank FC matrices in presence of dense noise (i.e., many of the FC edges are corrupted) than the generic PCP. We develop an efficient algorithm via linearized alternating direction method of multipliers (ADMM) to solve the proposed fused PCP program which admits no closed-form solution. We evaluated the performance of the proposed algorithm for L+S recovery via simulations. Application to naturalistic fMRI data reveals dynamic FC patterns that are reliably related to processing of continuous auditory events during movie watching.

The contributions of this work are as follows:
(1) This is the first work to propose a L+S decomposition method for naturalistic fMRI, which is is model-free and data-driven in detecting stimulus-induced dynamic FC.
(2) The proposed L+S learning algorithm offers a novel way to isolate low-rank parts of correlated multi-subject FC from idiosyncratic components over time. It exploits shared neuronal activity across subjects in terms of the network connectivity directly rather than the low-level BOLD responses in ISFC. This is equivalent to denoising the FC networks instead of raw fMRI signals via ISC, and thus extracts better FC structure information. By utilizing cross-subject reliability of FC measures captured in the low-rank part, we show that it provides a better mapping to naturalistic stimuli than the ISFC. Furthermore, our method advances a more general analysis framework applicable to other FC measures beyond the Pearson correlation in ISFC. (3) Unlike ISFC which computes a group-level FC profile related only to the stimuli, the proposed method recovers both the stimulus-related and unrelated FC patterns. This allows us to quantify confounding influence on the stimulus-induced FC, including potentially interesting stimulus-independent sources of neural dynamics, such as intrinsic processes, attentional variations and other individual differences. (4) While extracting a shared connectivity structure, our method also retains subject-specific FC patterns unavailable from ISFC, which is useful for studying individual differences in dynamic FC during naturalistic stimulation.

\vspace{-0.1in}
\section{Low-Rank and Sparse Matrix Decomposition for Dynamic Functional Connectivity}

\subsection{Proposed Model}

Suppose we observe a set of weighted connectivity matrices $\{\boldsymbol{\Sigma}_{ti}, t=1, \ldots, T, i=1, \ldots, M\}$ for multi-subject, time-dependent functional brain networks with same set of $N$ nodes, where elements in $\boldsymbol{\Sigma}_{ti} \in \mathbb{R}^{N \times N}$ are measures of pairwise interactions between nodes in the network at time point $t$ for $i$th subject. $T$ is the number of time points and $M$ is the number of subjects. Given the observed $\{\boldsymbol{\Sigma}_{ti}\}$, we aim to separate the task or stimulus-related brain FC dynamics from the background composed of intrinsic neural correlations, noise and other non-neuronal contributions. Our approach builds on the hypothesis that FC networks across different subjects admit a common underlying structure and exhibit similar or highly-correlated patterns, due to shared neuronal response among subjects when experiencing the same stimuli, e.g., watching the same movie in a naturalistic setting. On the other hand, the background connectivity networks are assumed to be relatively sparse compared to the stimulus-induced ones, and some confounding fluctuations such as motion-related artifacts occur only occasionally in certain subjects. These background activities are spontaneous, not time-locked to stimuli, and thus varying considerably across subjects.

In particular, we propose a decomposition of the time-dependent connectivity matrices $\boldsymbol{\Sigma}_{ti}$ as a superposition of a low-rank and a sparse component that represent respectively the correlated stimulus-induced connectivity and the subject-varying background activities. To formulate, let ${\bf z}_{ti} = \text{vec}(\boldsymbol{\Sigma}_{ti})$ be the vectorized version of $\boldsymbol{\Sigma}_{ti}$, and ${\bf Z}_t =  [{\bf z}_{t1}, \ldots, {\bf z}_{tM}] \in \mathbb{R}^{N^2 \times M}$ to denote the concatenation of the vectors of connectivity metrics over all subjects at time $t$. We consider a low-rank plus sparse (L+S) decomposition of the time-dependent matrices ${\bf Z}_t$
\begin{equation} \label{Eq:L+S}
{\bf Z}_t = {\bf L}_t + {\bf S}_t, \ \ \ \forall t=1, \ldots, T
\end{equation}
where ${\bf L}_t = [{\bf l}_{t1}, \ldots, {\bf l}_{tM}]$ is the low-rank matrix with $\text{rank}({\bf L}_t) = r_t < \min(N^2,M)$, ${\bf S}_t = [{\bf s}_{t1}, \ldots, {\bf s}_{tM}]$ is a sparse matrix with a fraction $s$ of non-zero entries, and ${\bf l}_{ti} \in \mathbb{R}^{N^2} $, ${\bf s}_{ti} \in \mathbb{R}^{N^2} $ are respectively the column vectors of stimulus-related and background connectivity metrics for the $i$-th subject at time $t$. The stimulus-induced connectivity that are correlated between subjects at each time $t$ can be well-approximated by the columns of ${\bf L}_t$ that lie on a low-dimensional subspace spanned by common bases. For example, rank one $r_t=1$ yields a constant connectivity pattern over all subjects. The columns of ${\bf S}_t$ captures the subject-specific background activities. Unlike the small Gaussian noise assumption, entries of ${\bf S}_t$ may have arbitrarily large magnitude with unknown sparse support, and thus capturing random, gross corruptions from the background events on the stimulus-related information in functional connectivity.

\vspace{-0.1in}

\subsection{L+S Decomposition}
\subsubsection{Principal Component Pursuit}
To recover the sequence of low-rank matrices $\{{\bf L}_t\}$ of stimulus-related connectivity from the corrupted observations $\{{\bf Z}_t\}$ in (\ref{Eq:L+S}), we consider solving the following optimization problem
\begin{align}
\min_{\{{\bf L}_t\},\{{\bf S}_t\}} & \sum_{t=1}^T \left\|{\bf L}_t\right\|_{*} + \lambda \sum_{t=1}^T \left\|{\bf S}_t\right\|_{1} \notag \\ s.t. & \ \ {\bf L}_t+{\bf S}_t = {\bf Z}_t, \ \ \forall t=1, \ldots, T \label{Eq:PCP}
\end{align}
where $\left\|{\bf L}_t\right\|_{*} = \sum_{i=1}^{r_t} \sigma_i({\bf L}_t)$ is the nuclear norm of matrix ${\bf L}_t$, i.e. the sum of all its singular values, $\left\|{\bf S}_t\right\|_{1} = \sum_{ij} |s_{tij}|$ is the $\ell_1$ norm of ${\bf S}_t$ and $\lambda > 0$ is a regularization parameter controlling the trade-off between the low-rank and sparse components. The nuclear norm induces sparsity of the vector of singular values, or equivalently for the matrix ${\bf L}_t$ to be low rank, while the $\ell_1$ penalty imposes element-wise sparsity of ${\bf S}_t$. Minimization of (\ref{Eq:PCP}) corresponds to applying the well-known PCP approach \cite{Candes2011} at each time point. Under the incoherence conditions on the row and column subspaces of ${\bf L}_t$, the PCP solution can perfectly recover the low-rank and sparse components, provided that the rank of ${\bf L}_t$ is not too large and ${\bf S}_t$ is reasonably sparse \cite{Candes2011,Chandrasekaran2011}. Many efficient and scalable algorithms such as the augmented Lagrange multipliers (ALM) method \cite{Lin2010} have been proposed to solve this convex optimization.

\subsubsection{Fused L+S Recovery}
The slowly-varying or possibly constant stimulus-related connectivity patterns across subjects may not be fully explained by a low-rank structure estimated in (\ref{Eq:PCP}). To better capture this inter-subject similarity in connectivity structure, we introduce a novel L+S decomposition by solving a fused version of PCP (fused PCP)
\begin{align}
\min_{\{{\bf L}_t\},\{{\bf S}_t\}} \sum_{t=1}^T \left\|{\bf L}_t\right\|_{*} & + \lambda_1 \sum_{t=1}^T \left\|{\bf S}_t\right\|_{1} \notag  \\ & + \lambda_2 \sum_{t=1}^T \sum_{i=2}^M \left\|{\bf l}_{ti} - {\bf l}_{t(i-1)} \right\|_{1} \notag \\ s.t. \ \  {\bf L}_t+{\bf S}_t & = {\bf Z}_t, \ \ \forall t=1, \ldots, T \label{Eq:FusedLS}
\end{align}
Here, the additional fused term $\sum_{i=2}^M \left\|{\bf l}_{ti} - {\bf l}_{t(i-1)} \right\|_{1}$, regularized by tuning parameter $\lambda_2>0$, encourages the stimulus-induced group homogeneity by penalizing the differences in the connectivity profiles ${\bf l}_{ti}$ between subjects. For some sufficiently large $\lambda_2$, the penalty shrinks inter-subject differences $|l_{tij} - l_{t(i-1)j}|$ in some connectivity coefficients $j$ to 0, resulting in similar (but not necessarily identical) connectivity patterns across subjects.
Clearly, the model (\ref{Eq:FusedLS}) includes the original L+S decomposition (\ref{Eq:PCP}) as a special case when $\lambda_2=0$.

Define a matrix ${\bf A} = {\bf D} \otimes {\bf I}_{N^2}$ where
\[
{\bf D} =
\left(
  \begin{array}{ccccc}
    -1 & 1 & 0 & \ldots & 0 \\
    0 & -1 & 1 & \ldots & 0 \\
    \vdots & \vdots  & \ddots & \ddots & \vdots \\
		0 & 0 & \ldots & -1 & 1 \\
  \end{array}
\right) \in \mathbb{R}^{(M-1) \times M}
\]
is first-order difference matrix and ${\bf I}_{N^2}$ is a $N^2 \times N^2$ identity matrix. We can reformulate the objective function in (\ref{Eq:FusedLS}) as
\begin{equation} \label{Eq:FusedLS-diff}
\sum_{t=1}^T \left\|{\bf L}_t\right\|_{*} + \lambda_1 \sum_{t=1}^T \left\|{\bf S}_t\right\|_{1} + \lambda_2 \sum_{t=1}^T \left\|{\bf A}\text{vec}({\bf L}_t)\right\|_{1}.
\end{equation}

\vspace{-0.1in}
\section{Optimization}

For the fused L+S decomposition, we develop an ADMM algorithm to solve the convex optimization problem (\ref{Eq:FusedLS}) which has no closed form solution. We introduce a set of surrogate variables $\boldsymbol{\alpha}_1, \ldots, \boldsymbol{\alpha}_T$ where $\boldsymbol{\alpha}_t = {\bf A}\text{vec}({\bf L}_t) \in \mathbb{R}^{N^2(M-1)}$. Then, the minimization of (\ref{Eq:FusedLS-diff}) is equivalent to
\begin{align}
\min_{\{{\bf L}_t\},\{{\bf S}_t\},\{\boldsymbol{\alpha}_t\}} \sum_{t=1}^T \left\|{\bf L}_t\right\|_{*} & + \lambda_1 \sum_{t=1}^T \left\|{\bf S}_t\right\|_{1} + \lambda_2 \sum_{t=1}^T \left\| \boldsymbol{\alpha}_t \right\|_{1} \notag \\ s.t. \ \  {\bf L}_t+{\bf S}_t = {\bf Z}_t, \ \ & {\bf A}\text{vec}({\bf L}_t) = \boldsymbol{\alpha}_t, \ \ \forall t=1, \ldots, T \label{Eq:FusedLS-alpha}
\end{align}
The augmented Lagrangian function of (\ref{Eq:FusedLS-alpha}) is
\begin{flalign}
L_{\mu}&(\{{\bf L}_t\},\{{\bf S}_t\},\{\boldsymbol{\alpha}_t\} ,\{\widetilde{\bf X}_t\},\{\widetilde{\bf Y}_t\}) \notag \\ & = \sum_{t=1}^T \left\|{\bf L}_t\right\|_{*} + \lambda_1 \sum_{t=1}^T \left\|{\bf S}_t\right\|_{1} + \lambda_2 \sum_{t=1}^T \left\| \boldsymbol{\alpha}_t \right\|_{1} \notag \\ & + \sum_{t=1}^T \left( \tr[\widetilde{\bf X}_t^T({\bf L}_t+{\bf S}_t-{\bf Z}_t)] + \widetilde{\bf Y}_t^T({\bf A}\text{vec}({\bf L}_t)-\boldsymbol{\alpha}_t)\right) \notag \\ & + \frac{\mu}{2} \sum_{t=1}^T \left( \left\| {\bf L}_t+{\bf S}_t-{\bf Z}_t \right\|_{F}^2 + \left\| {\bf A}\text{vec}({\bf L}_t)-\boldsymbol{\alpha}_t \right\|_{2}^2 \right) \label{Eq:Lagr}
\end{flalign}
where $\{\widetilde{\bf X}_t\},\{\widetilde{\bf Y}_t\}$ are Lagrange multipliers, $\mu>0$ is a pre-specified step-size parameter, $\left\|\cdot\right\|_2$ denotes the $\ell_2$ norm and $\left\|\cdot\right\|_{F}$ is the Frobenius norm.
In a rescaled form, (\ref{Eq:Lagr}) can be written as
\begin{flalign}
L_{\mu}&(\{{\bf L}_t\},\{{\bf S}_t\},\{\boldsymbol{\alpha}_t\},\{{\bf X}_t\},\{{\bf Y}_t\}) \notag \\ & = \sum_{t=1}^T \left\|{\bf L}_t\right\|_{*} + \lambda_1 \sum_{t=1}^T \left\|{\bf S}_t\right\|_{1} + \lambda_2 \sum_{t=1}^T \left\| \boldsymbol{\alpha}_t \right\|_{1} \notag \\ & + \frac{\mu}{2} \sum_{t=1}^T \left( \left\|{\bf L}_t+{\bf S}_t-{\bf Z}_t+{\bf X}_t \right\|_{F}^2 \right. \notag \\ & + \left. \left\| {\bf A}\text{vec}({\bf L}_t)-\boldsymbol{\alpha}_t+{\bf Y}_t\right\|_{2}^2  - \left\|{\bf X}_t\right\|_{F}^2 - \left\|{\bf Y}_t\right\|_{2}^2 \right) \label{Eq:Lagr-sc}
\end{flalign}
where ${\bf X}_t = \mu^{-1}\widetilde{\bf X}_t$ and ${\bf Y}_t = \mu^{-1}\widetilde{\bf Y}_t$ are rescaled Lagrange multipliers.

We can obtain the parameter estimates by minimizing the augmented Lagrangian via ADMM algorithm \cite{Boyd2011}, which alternates between minimizing (\ref{Eq:Lagr-sc}) with respect to each set of primal variables ${\bf L}_t,{\bf S}_t,\boldsymbol{\alpha}_t$ sequentially, and the updating of the dual variables ${\bf X}_t,{\bf Y}_t$. Let initialize the parameters by $({\bf L}_t^0,{\bf S}_t^0,\boldsymbol{\alpha}_t^0,{\bf X}_t^0,{\bf Y}_t^0) = (\bf{0},\bf{0},\bf{0},\bf{0},\bf{0})$ for $t=1, \ldots, T$. The proposed ADMM algorithm solves the following subproblems iteratively until convergence, with parameter updates at each iteration $k$ given by
\begin{align}
\{{\bf L}_t^k\} & = \argmin_{\{{\bf L}_t\}} L_{\mu}(\{{\bf L}_t\},\{{\bf S}_t^{k-1}\},\{\boldsymbol{\alpha}_t^{k-1}\},\{{\bf X}_t^{k-1}\},\{{\bf Y}_t^{k-1}\}) \label{Eq:ADMM-1} \\
\{{\bf S}_t^k\} & = \argmin_{\{{\bf S}_t\}} L_{\mu}(\{{\bf L}_t^k\},\{{\bf S}_t\},\{\boldsymbol{\alpha}_t^{k-1}\},\{{\bf X}_t^{k-1}\},\{{\bf Y}_t^{k-1}\}) \label{Eq:ADMM-2} \\
\{\boldsymbol{\alpha}_t^k\} & = \argmin_{\{\boldsymbol{\alpha}_t\}} L_{\mu}(\{{\bf L}_t^k\},\{{\bf S}_t^k\},\{\boldsymbol{\alpha}_t\},\{{\bf X}_t^{k-1}\},\{{\bf Y}_t^{k-1}\}) \label{Eq:ADMM-3} \\
{\bf X}_t^k & = {\bf X}_t^{k-1} + {\bf L}_t^k + {\bf S}_t^k - {\bf Z}_t, \ \ \forall t=1, \ldots, T \label{Eq:ADMM-4} \\
{\bf Y}_t^k & = {\bf Y}_t^{k-1} + {\bf A}\text{vec}({\bf L}_t^k)-\boldsymbol{\alpha}_t^k, \ \ \forall t=1, \ldots, T \label{Eq:ADMM-5}
\end{align}
Next we derive the solution for each subproblem.

\subsubsection{Update ${\bf L}_t$} To update $\{{\bf L}_t^k\}$, the subproblem (\ref{Eq:ADMM-1}) can be separated into minimizations with respect to ${\bf L}_t$'s at individual time points (similarly for subproblems (\ref{Eq:ADMM-2})-(\ref{Eq:ADMM-3})), where each ${\bf L}_t$ is updated independently
\begin{align}
{\bf L}_t^k & = \argmin_{{\bf L}_t} \left\|{\bf L}_t\right\|_{*} + \frac{\mu}{2}\left\|{\bf L}_t+{\bf S}_t^{k-1}-{\bf Z}_t+{\bf X}_t^{k-1} \right\|_{F}^2 \notag \\ & + \frac{\mu}{2} \left\| {\bf A}\text{vec}({\bf L}_t)-\boldsymbol{\alpha}_t^{k-1}+{\bf Y}_t^{k-1}\right\|_{2}^2. \label{Eq:updateL}
\end{align}
Due to the non-identity matrix ${\bf A}$, (\ref{Eq:updateL}) does not have a closed-form solution. Motivated by the linearized ADMM method in \cite{Wang2012} recently applied to lasso regression \cite{Li2014,Li2020}, we linearize the quadratic term $\frac{1}{2} \left\| {\bf A}\text{vec}({\bf L}_t)-\boldsymbol{\alpha}_t^{k-1}+{\bf Y}_t^{k-1}\right\|_{2}^2$ in (\ref{Eq:updateL}) as
\begin{align}
\left({\bf A}^T\left({\bf A}\text{vec}({\bf L}_t^{k-1})-\boldsymbol{\alpha}_t^{k-1}+{\bf Y}_t^{k-1}\right)\right)^T & \left(\text{vec}({\bf L}_t)-\text{vec}({\bf L}_t^{k-1}) \right) \notag \\ + \frac{\nu}{2} \left\| \text{vec}({\bf L}_t)-\text{vec}({\bf L}_t^{k-1}) \right\|_{2}^2 & \label{Eq:Linearterm}
\end{align}
where the parameter $\nu>0$ controls the proximity to ${\bf L}_t^{k-1}$.
Substituting (\ref{Eq:Linearterm}) into (\ref{Eq:updateL}) and after some algebra, we obtain the following approximate subproblem for (\ref{Eq:updateL})
\begin{align}
{\bf L}_t^k & = \argmin_{{\bf L}_t} \frac{1}{\mu} \left\|{\bf L}_t\right\|_{*} + \frac{1}{2} \left\|{\bf L}_t+{\bf S}_t^{k-1}-{\bf Z}_t+{\bf X}_t^{k-1} \right\|_{F}^2 \notag \\ & + \frac{\nu}{2} \left\|{\bf L}_t - {\bf C}_t \right\|_{F}^2 \label{Eq:updateL-linear}
\end{align}
where ${\bf C}_t \in \mathbb{R}^{N^2 \times M}$ is a matrix such that $\text{vec}({\bf C}_t) = {\bf c}_t$ with ${\bf c}_t = \text{vec}({\bf L}_t^{k-1}) - {\bf A}^T\left({\bf A}\text{vec}({\bf L}_t^{k-1})-\boldsymbol{\alpha}_t^{k-1}+{\bf Y}_t^{k-1}\right)/\nu$.
We show that (\ref{Eq:updateL-linear}) can be solved via the singular value thresholding (SVT) \cite{Cai2010}. More precisely, let $\mathcal{S}_{\tau}({\bf M})$ be the element-wise soft-thresholding (shrinkage) operator for a matrix ${\bf M}$ at level $\tau>0$, $(\mathcal{S}_{\tau}({\bf M}))_{ij} = \sign(m_{ij})\max(|m_{ij}-\tau|,0)$. We also denote by $\mathcal{D}_{\tau}({\bf M})$ the singular value thresholding operator given by $\mathcal{D}_\tau({\bf M}) = {\bf U}\mathcal{S}_\tau(\boldsymbol{\Sigma}){\bf V}^T$ where ${\bf M} = {\bf U}\boldsymbol{\Sigma}{\bf V}^T$ is the singular value decomposition (SVD) of ${\bf M}$. Then, ${\bf L}_t^k$ admits an explicit solution
\begin{equation}
{\bf L}_t^k = \mathcal{D}_{1/\mu(\nu+1)}({\bf Q}_t), \ \ \ {\bf Q}_t = \frac{1}{\nu+1}({\bf Z}_t - {\bf S}_t^{k-1} + \nu{\bf C}_t - {\bf X}_t^{k-1}) \label{Eq:L-sol}
\end{equation}
The derivation of (\ref{Eq:L-sol}) is given in Supplementary Section I. Note that when $\nu \rightarrow 0$, (\ref{Eq:L-sol}) reduces to the solution of the ALM algorithm for the low-rank component \cite{Lin2010} in the original L+S decomposition problem (\ref{Eq:PCP}).

\subsubsection[Update ${S}_t$]{Update ${\bf S}_t$} The minimization in subproblem (\ref{Eq:ADMM-2}) with respect to ${\bf S}_t$ at each time point is equivalent to
\begin{equation}
{\bf S}_t^k = \argmin_{{\bf S}_t} \frac{\mu}{2}\left\|{\bf L}^k_t+{\bf S}_t-{\bf Z}_t+{\bf X}_t^{k-1} \right\|_{F}^2 + \lambda_1 \left\|{\bf S}_t\right\|_{1} \notag
\end{equation}
We can get a closed-form solution by applying the element-wise soft-thresholding operator
\begin{equation}
{\bf S}_t^k = \mathcal{S}_{\lambda_1/\mu}({\bf Z}_t-{\bf L}_t^k-{\bf X}_t^{k-1}) \label{Eq:S-sol}.
\end{equation}

\subsubsection[Update ${\alpha}_t$]{Update $\boldsymbol{\alpha}_t$} Similarly, the subproblem (\ref{Eq:ADMM-3}) with respect to each $\boldsymbol{\alpha}_t$ is
\begin{equation}
\boldsymbol{\alpha}_t^k = \argmin_{\boldsymbol{\alpha}_t} \frac{\mu}{2} \left\| {\bf A}\text{vec}({\bf L}_t^k)-\boldsymbol{\alpha}_t+{\bf Y}_t^{k-1}\right\|_{2}^2 + \lambda_1 \left\|\boldsymbol{\alpha}_t\right\|_{1} \notag
\label{Eq:updatealpha}
\end{equation}
where the solution is also a soft-thresholding operation
\begin{equation}
\boldsymbol{\alpha}_t^k = \mathcal{S}_{\lambda_2/\mu}({\bf A}\text{vec}({\bf L}_t^k)+{\bf Y}_t^{k-1}) \label{Eq:alpha-sol}.
\end{equation}

We chose the parameter $\lambda_1=1/\sqrt{\max(N^2,M)}$, as suggested by \cite{Candes2011} for the PCP. The tuning parameter $\lambda_2$ can be selected using cross-validation. To establish the convergence of the linearized ADMM algorithm, the parameter $\nu$ is required to satisfy $\nu > \rho({\bf A}^T{\bf A})$ where $\rho(.)$ denotes the spectral radius \cite{Li2020}. We set $\nu = 1.5\rho({\bf A}^T{\bf A})$.
We set the stopping criterion as
\begin{equation}
\frac{\left\|({\bf L}_t^k,{\bf S}_t^k) - ({\bf L}_t^{k-1},{\bf S}_t^{k-1})\right\|_{F}}{\max\left\{ \left\|({\bf L}_t^{k-1},{\bf S}_t^{k-1})\right\|_{F},1 \right\}} \leq 10^{-6}, \ \forall t=1, \ldots, T \notag
\end{equation}
or the maximum iteration number of 5000 is reached. The proposed ADMM algorithm for solving the fused L+S decomposition is summarized in Algorithm 1.

\begin{algorithm}[t]
\caption{ADMM for Fused L+S Decomposition}
\begin{algorithmic}[1]
\renewcommand{\algorithmicrequire}{\textbf{Input:}}
\renewcommand{\algorithmicensure}{\textbf{Output:}}
\REQUIRE A series of vectorized multi-subject dynamic connectivity metrics ${\bf Z}_1, \ldots, {\bf Z}_T$.
\renewcommand{\algorithmicrequire}{\textbf{Parameters:}}
\REQUIRE $\lambda_1=1/\sqrt{\max(N^2,M)}$, $\lambda_2>0$, $\mu = 1/\left\|{\bf Z}_t\right\|_{2}$, $\nu > \rho({\bf A}^T{\bf A})$ where $\rho(.)$ denotes the spectral radius.
\renewcommand{\algorithmicrequire}{\textbf{Initialize:}}
\REQUIRE $({\bf L}_t^0,{\bf S}_t^0,\boldsymbol{\alpha}_t^0,{\bf X}_t^0,{\bf Y}_t^0) = (\bf{0},\bf{0},\bf{0},\bf{0},\bf{0})$, $\forall t=1, \ldots, T$.
\FOR {$t = 1:T$}
	\WHILE {not converge}
		\STATE Compute ${\bf C}_t$: $\text{vec}({\bf C}_t) = {\bf c}_t$ with ${\bf c}_t = \text{vec}({\bf L}_t^{k-1}) - {\bf A}^T\left({\bf A}\text{vec}({\bf L}_t^{k-1})-\boldsymbol{\alpha}_t^{k-1}+{\bf Y}_t^{k-1}\right)/\nu$
		\STATE Compute ${\bf Q}_t = \frac{1}{\nu+1}({\bf Z}_t - {\bf S}_t^{k-1} + \nu{\bf C}_t - {\bf X}_t^{k-1})$
		\STATE Update ${\bf L}_t^k = \mathcal{D}_{1/\mu(\nu+1)} \left({\bf Q}_t\right)$
		\STATE Update ${\bf S}_t^k = \mathcal{S}_{\lambda_1/\mu}({\bf Z}_t-{\bf L}_t^k-{\bf X}_t^{k-1})$
		\STATE Update $\boldsymbol{\alpha}_t^k = \mathcal{S}_{\lambda_2/\mu}({\bf A}\text{vec}({\bf L}_t^k)+{\bf Y}_t^{k-1})$
		\STATE Update ${\bf X}_t^k = {\bf X}_t^{k-1} + {\bf L}_t^k + {\bf S}_t^k - {\bf Z}_t$
		\STATE Update ${\bf Y}_t^k = {\bf Y}_t^{k-1} + {\bf A}\text{vec}({\bf L}_t^k)-\boldsymbol{\alpha}_t^k$
	\ENDWHILE
\ENDFOR
\ENSURE $\widehat{\bf L}_1, \ldots, \widehat{\bf L}_T$ and $\widehat{\bf S}_1, \ldots, \widehat{\bf S}_T$.
\end{algorithmic}
\end{algorithm}

\vspace{-0.15in}
\section{Simulation}

We evaluate the performance of the proposed fused L+S decomposition method via simulation studies. The goal is to assess the performance of L+S algorithms in separating the common structure from individual-specific background components in multi-subject FC networks. Here, we focus on the static snapshot of dynamic networks at a particular time point, and drop the index $t$ for notational brevity. We generate the data matrix of concatenated connectivity metrics of weighted, undirected networks for $M$ subjects as follows
\begin{equation}
{\bf Z} = {\bf B}\boldsymbol{\beta} + {\bf S}
\end{equation}
where ${\bf Z} =  [{\bf z}_1, \ldots, {\bf z}_M]$, ${\bf S} =  [{\bf s}_1, \ldots, {\bf s}_M]$ with ${\bf s}_i = \text{vec}\left(\boldsymbol{\Sigma}_i^{s}\right)$, ${\bf B} = [{\bf b}_1, \ldots, {\bf b}_r] \in \mathbb{R}^{N^2 \times r}$ whose columns correspond to a set of $r$ vectorized basis connectivity matrices ${\bf b}_j = \text{vec}\left(\boldsymbol{\Sigma}_j^{b}\right)$ that are shared across subjects, and $\boldsymbol{\beta} = [\boldsymbol{\beta}_1, \ldots, \boldsymbol{\beta}_M]$ with $\boldsymbol{\beta}_i \in \mathbb{R}^r$ being a vector of mixing weights for subject $i$. To emulate the modular structure of brain networks \cite{Sporns2016} induced by stimuli or tasks, we generate each basis connectivity matrix $\boldsymbol{\Sigma}_j^{b}$ from a stochastic block model (SBM) \cite{Pavlovic2014} with two equal-sized communities with intra- and inter-community edge probabilities of 0.95 and 0.2, respectively. The edge strengths are randomly drawn from the uniform distribution $U[-1,1]$. The subject-dependent mixing weights $\boldsymbol{\beta}_i$ are clustered according to two subgroups, which are independently sampled from normal distributions $\boldsymbol{\beta}_i \sim N(0.5{\bf I},\epsilon{\bf I})$ for $i=1, \ldots, M/2$ and $\boldsymbol{\beta}_i \sim N({\bf 0},\epsilon{\bf I})$ for $i={M/2+1}, \ldots, M$ to represent the strong and the weak response to the stimulus in the two subgroups, respectively. We set $\epsilon = 0.005$ such that the variability within each subgroup is small. The resulting matrix ${\bf L} = {\bf B}\boldsymbol{\beta}$ has rank $r$, and the connectivity profiles in the columns of ${\bf L}$ are similar within each subgroup. The background components of individual subjects $\boldsymbol{\Sigma}_i^{s}$ are generated independently as sparse symmetric matrices with support set uniformly chosen at random at a constant sparsity level $s$ (fraction of non-zero entries), and whose non-zero entries are i.i.d. samples from $U[-5,5]$ indicating large magnitude of noise in ${\bf Z}$. The simulations were replicated 100 times.

We compare the performance of the fused L+S decomposition with the original version of PCP in recovering ${\bf L}$ and ${\bf S}$ from the observed matrix ${\bf Z}$. For the fused L+S, we select the optimal parameter $\lambda_2$ over a large grid of values, which gives the best recovery performance on an independently generated validation set. To measure the accuracy of recovery, we computed the root mean squared error (RMSE) between the ground-truth and estimated low-rank and sparse components, defined by $\text{RMSE-L} = {\|\widehat{\bf L} - {\bf L}\|_{F}}/\left\|{\bf L}\right\|_{F}$ and $\text{RMSE-S} = {\|\widehat{\bf S} - {\bf S}\|_{F}}/\left\|{\bf S}\right\|_{F}$.

\begin{figure*}[!t]
\centering
\captionsetup[subfigure]{labelformat=empty}
	\begin{minipage}[b]{0.3\linewidth}
		\centering
		\subfigure{\includegraphics[width=0.8\linewidth,keepaspectratio]{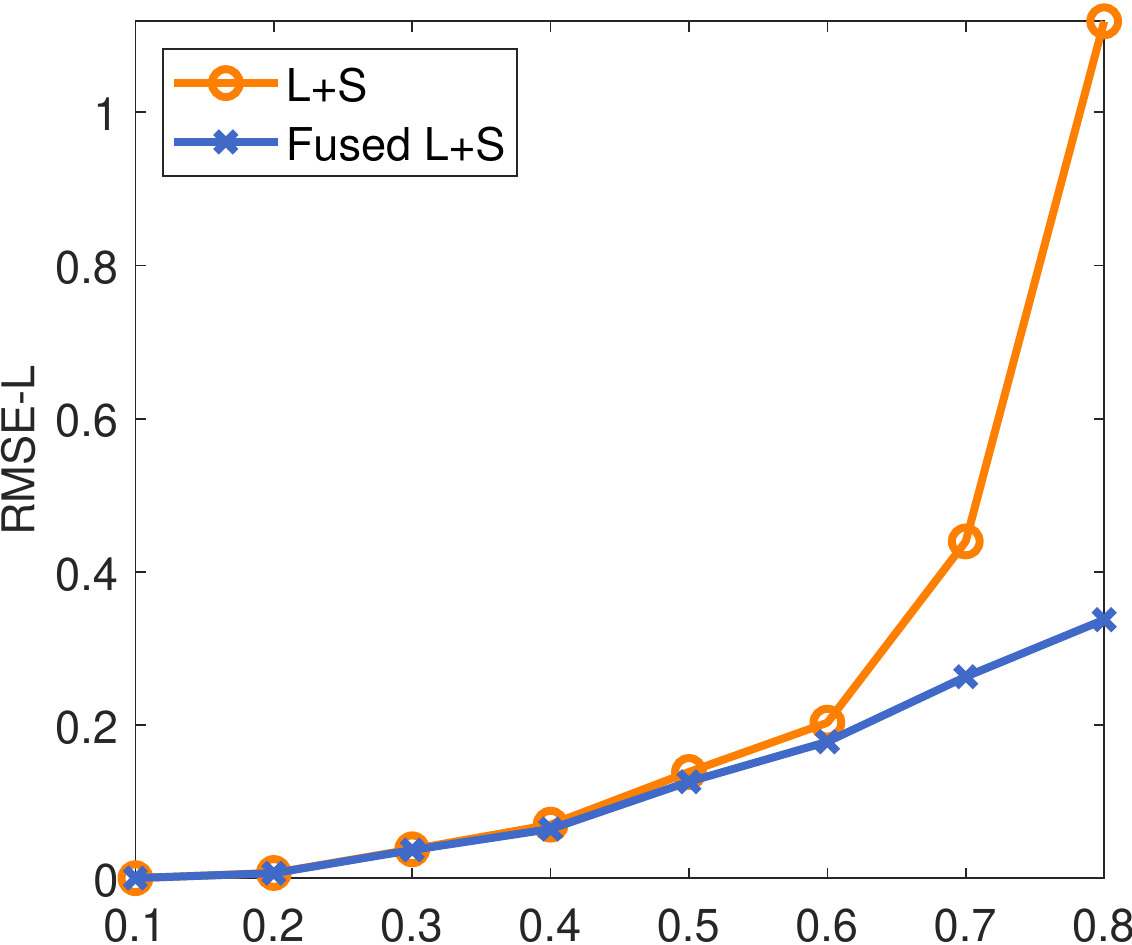}}
	\end{minipage}
	\hspace{-0.85 cm}
	\begin{minipage}[b]{0.3\linewidth}
		\centering
		\subfigure{\includegraphics[width=0.8\linewidth,keepaspectratio]{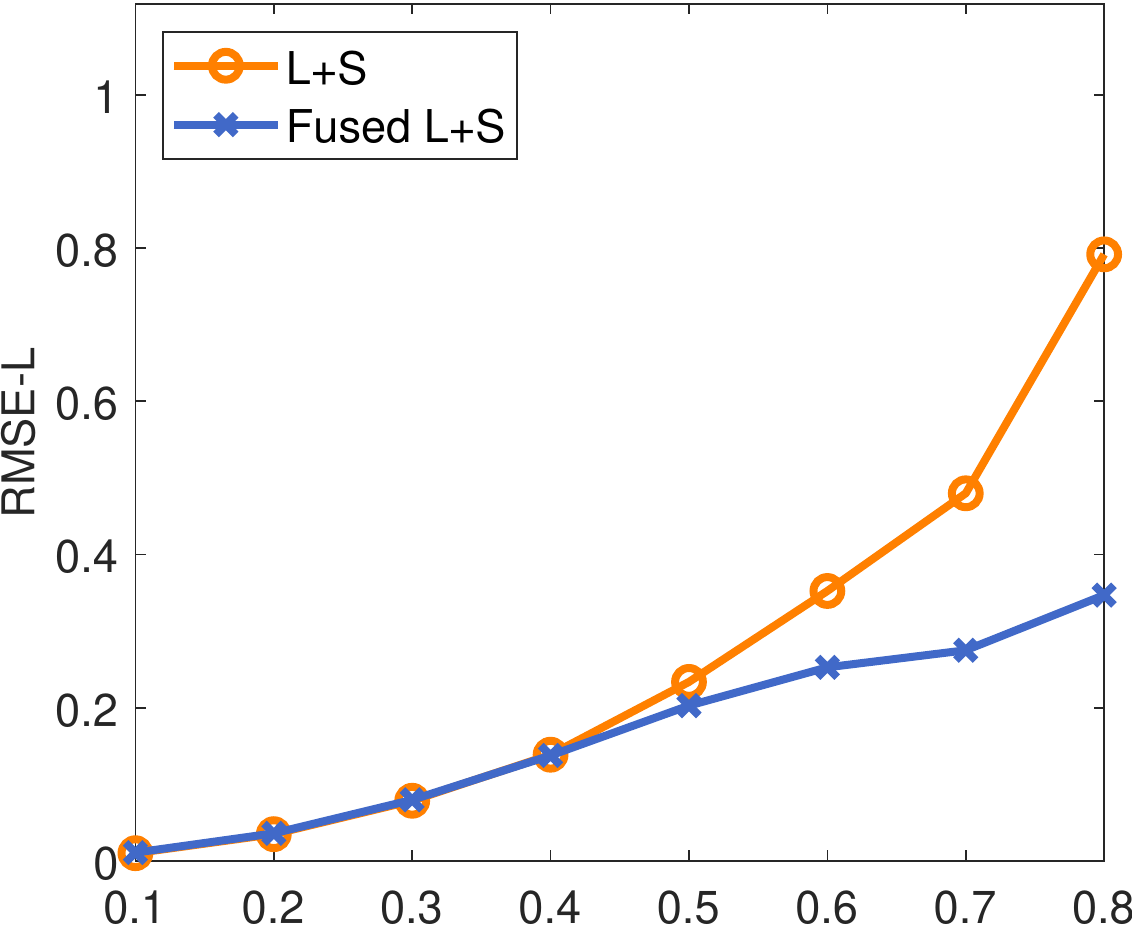}}
	\end{minipage}
	\hspace{-0.85 cm}
	\begin{minipage}[b]{0.3\linewidth}
		\centering
		\subfigure{\includegraphics[width=0.8\linewidth,keepaspectratio]{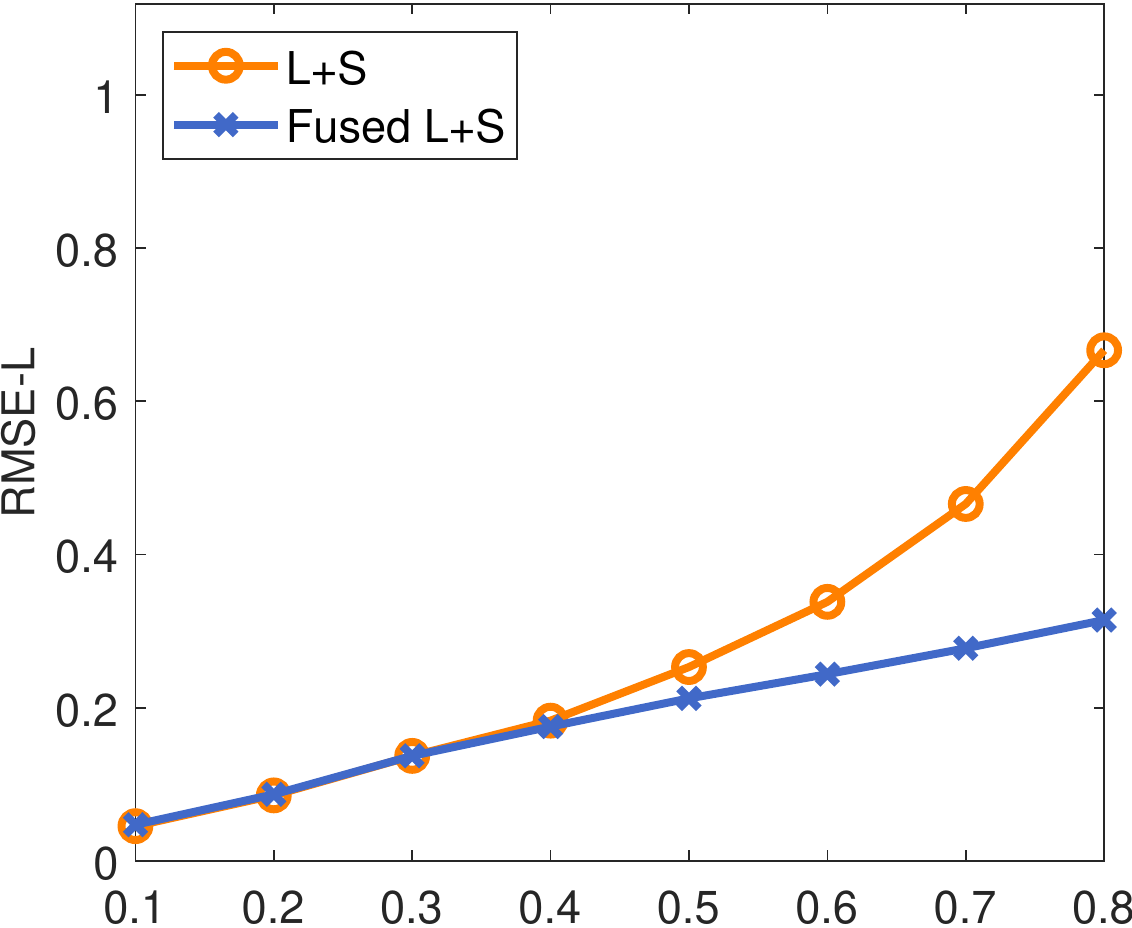}}
	\end{minipage}

	\vspace{0.3 cm}
		\begin{minipage}[b]{0.3\linewidth}
		\centering
		\subfigure[(a) $r=1$]{\includegraphics[width=0.8\linewidth,keepaspectratio]{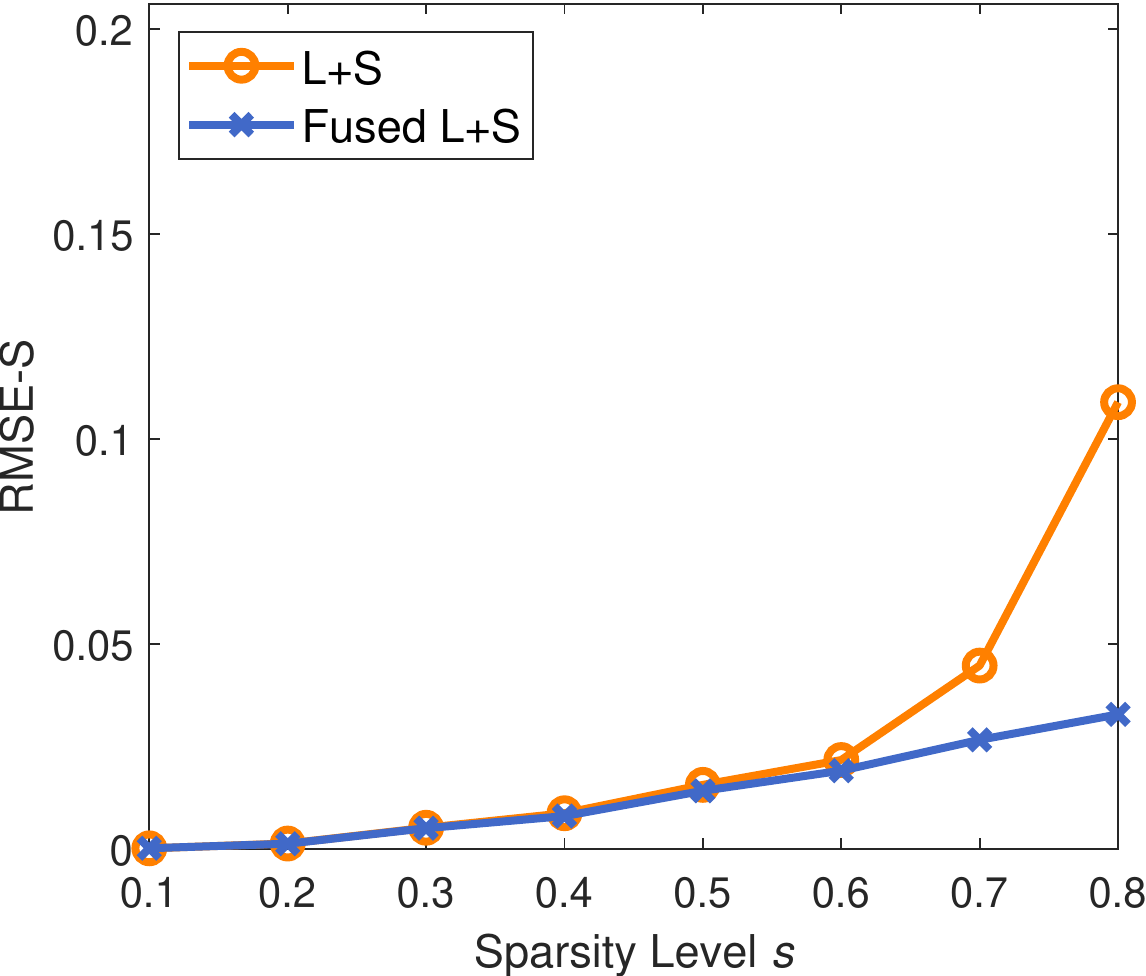}}
	\end{minipage}
	\hspace{-0.85 cm}
	\begin{minipage}[b]{0.3\linewidth}
		\centering
		\subfigure[(b) $r=5$]{\includegraphics[width=0.8\linewidth,keepaspectratio]{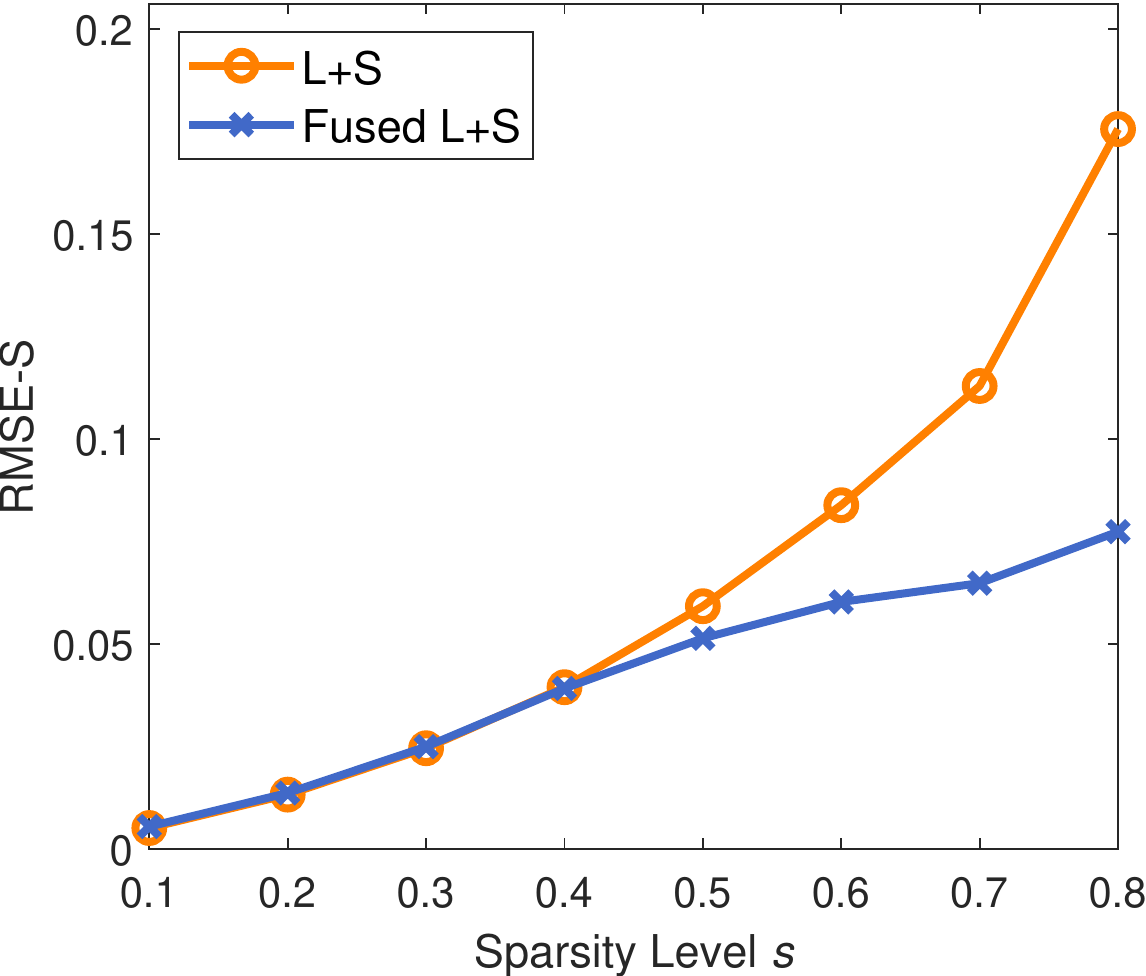}}
	\end{minipage}
	\hspace{-0.85 cm}
	\begin{minipage}[b]{0.3\linewidth}
		\centering
		\subfigure[(c) $r=10$]{\includegraphics[width=0.8\linewidth,keepaspectratio]{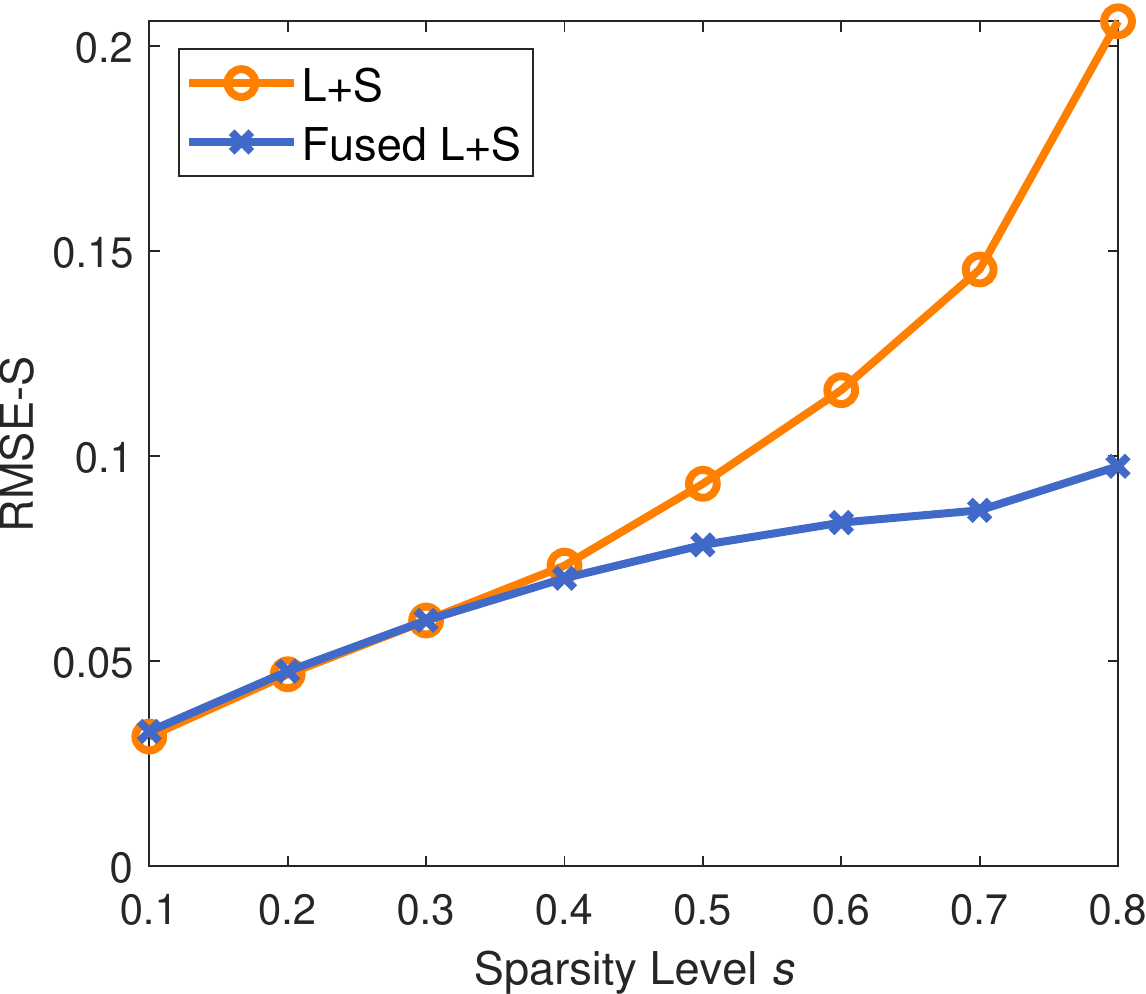}}
	\end{minipage}
\caption{Performance comparison of original and fused L+S decomposition algorithms in recovering low-rank and sparse structures from simulated multi-subject connectivity matrices. RMSEs of estimated low-rank (top) and sparse (bottom) components against increasing level of sparsity $s$ for different ranks $r$. Results are averages over 100 replications.}
\label{Fig:EffRs}
\vspace{-0.15in}
\end{figure*}

\begin{figure}[!t]
	\begin{minipage}[t]{1\linewidth}
		\centering
		\includegraphics[width=0.9\linewidth,keepaspectratio]{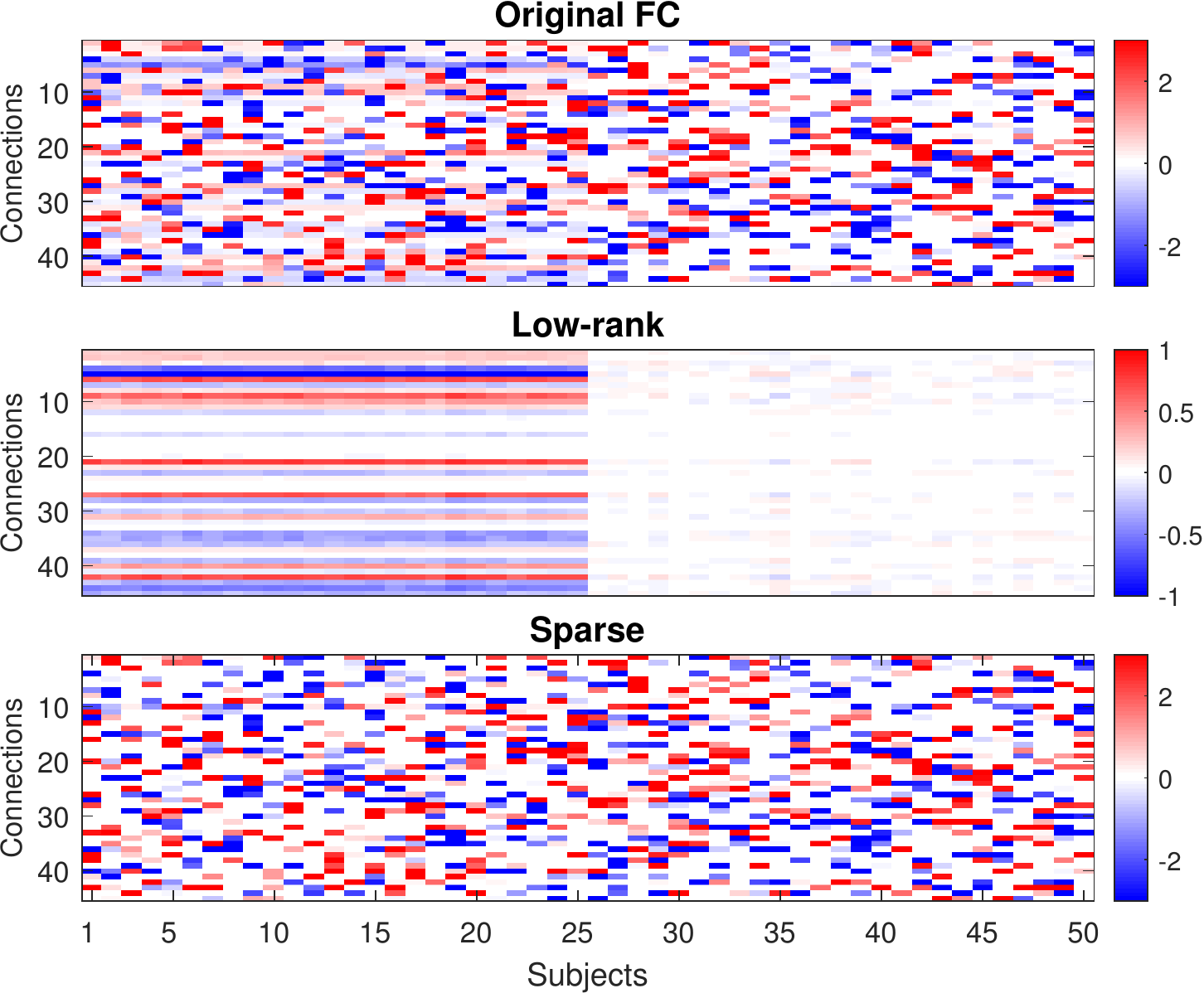}
	\end{minipage}
	\vspace{-0.25in}
\caption{L+S recovery of synthetic multi-subject FC networks. One realization of simulated matrix ${\bf Z}$ (top) whose columns are vectorized upper-triangular part of undirected FC matrices with $N=10$ nodes for $M=50$ subjects. The dimension is $N(N-1)/2 \times M = 45 \times 50$. Low-rank $\widehat{\bf L}$ (middle) and sparse $\widehat{\bf S}$ (bottom) components recovered from ${\bf Z}$ using the fused L+S method.}
\label{Fig:one-sim-LS}
	\vspace{-0.15in}
\end{figure}

We investigate robustness of the methods to increasing rank $r$ and sparsity level $s$ of the ground-truth components. Fig.\ref{Fig:EffRs} plots the average RMSEs over 100 replications as a function of $s$ (varied from 0.1 to 0.8 with an increment of 0.1) for $r=1,5,10$ with fixed $N=10$ and $M=50$. As expected, estimation errors of both methods for the low-rank and sparse components increase with $s$ and $r$, generally. The larger fraction of corrupted entries and the rank of the data matrix render the estimated $\widehat{\bf L}$ and $\widehat{\bf S}$ more deviated from the ground-truth, in consistency with theoretical results in previous studies \cite{Candes2011}. Both methods perform comparably when $s$ is small. However, as $s$ increases, the fused L+S estimator clearly outperforms the original version with substantially lower errors with a slower growth rate. This suggests the robustness of the fused L+S method under the presence of severe dense noise. It provides a better recovery of the low-rank matrix of shared connectivity patterns even if almost all of its entries are grossly corrupted by the background components, as evidenced by the relatively stable errors when $s=0.7$ and $s=0.8$. This is due to the additional fused penalty term which smooths out the individual-specific random background noise, when leveraging on the inter-subject similarity of the stimulus-induced connectivity profiles.

Fig.\ref{Fig:one-sim-LS} shows fused L+S decomposition of a simulated multi-subject connectivity matrix ${\bf Z}$ with underlying rank $r=5$ and a fraction $s=0.5$ of corrupted edges in each subject. The estimate $\widehat{\bf L}$ successfully recovers the correlated connectivity patterns across subjects with distinct sub-group shared responses from the highly-corrupted ${\bf Z}$, while $\widehat{\bf S}$ filters out  subject-specific background noise without prior information of the locations of the corrupted edges (the support of ${\bf S}$).

\begin{figure*}[!ht]
\centering
\includegraphics[width=0.9\linewidth,keepaspectratio]{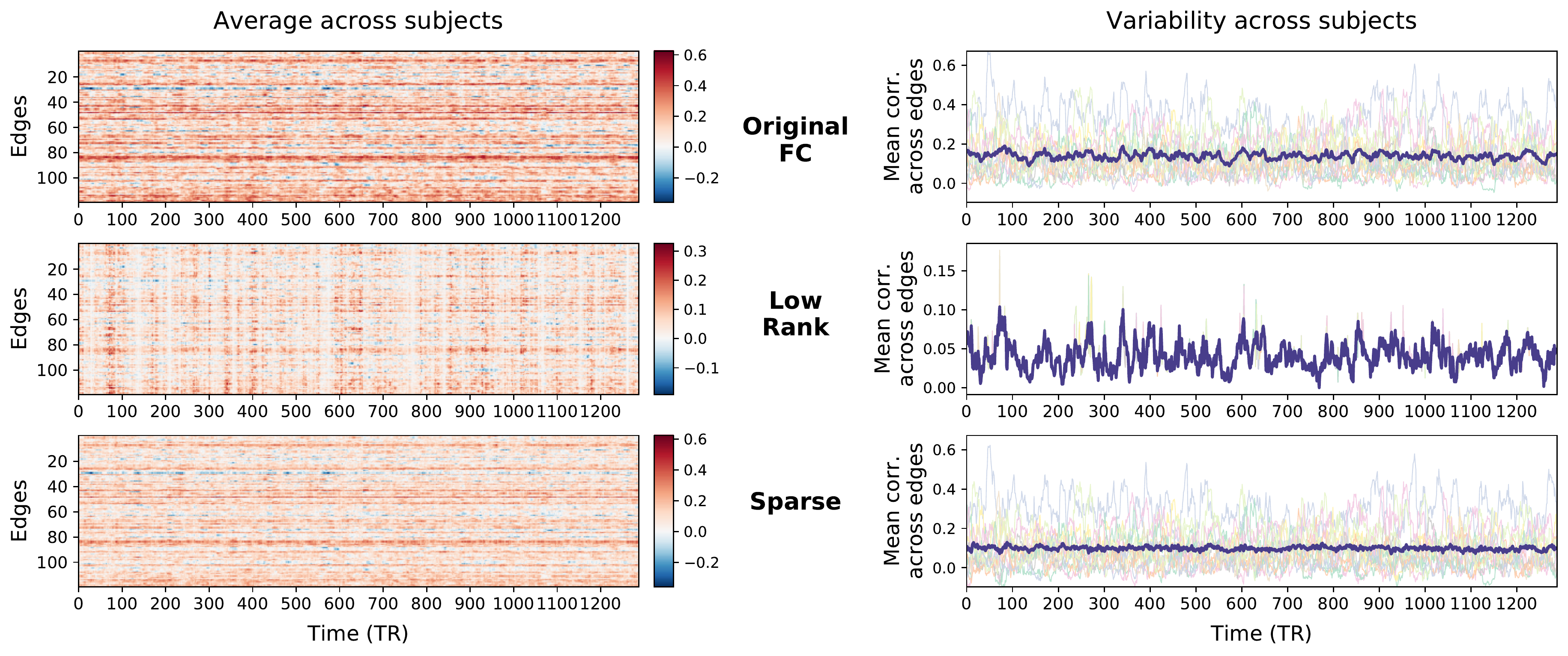}
\vspace{-0.1in}
	\caption{Fused L+S decomposition of time-resolved, multi-subject FC from movie-watching fMRI. (Left) Original dynamic pairwise correlations between 16 ROIs and the estimated low-rank and sparse components, averaged over all subjects. (Right) Mean dynamic correlations across all edges and the estimated low-rank and sparse components. The group-average (blue lines) and individual (thin lines) time courses depict distributions of dynamic correlations over subjects. The time-varying low-rank components are temporally locked across subjects, suggesting potential shared response to stimulus sequence in the movie.}  \label{Fig:tv-analysis}
\end{figure*}

\section{Analysis of Movie-Watching fMRI}

In this section, we applied the proposed fused L+S decomposition approach to separating the stimulus-related and background components in time-varying FC from fMRI data of a group of subjects exposed to complex naturalistic stimulation from movie viewing. We tested the established hypothesis that, as speech becomes more difficult to process, e.g., as through speech-in-noise, somatosensory, motor, and auditory systems are engaged more to facilitate processing \cite{Skipper2017}.

\vspace{-0.1in}
\subsection{Data Acquisition \& Stimuli}
Thirteen subjects watched a television show during which 3 Tesla fMRI data were collected at the Joan \& Sanford I. Weill Medical College of Cornell University by Dr Skipper (GE Medical Systems, Milwaukee, WI; TR = 1.5 s; see \cite{Skipper2017b} for full details). The show, \textit{Are you smarter than a 5th grader?}, was 32min 24sec long. Among other features, audio events were annotated at a 25ms resolution by a trained coder. These were grouped into five general categories including \textit{noise}, \textit{music}, \textit{speech} (speech that occurred in silence), \textit{sp+noise} (speech that co-occurred with noise, including background talking, clapping, music etc.) and \textit{silence}. The audio annotations were down-sampled to match the resolution of fMRI data by selecting the auditory category that occurred most frequently within each 1.5 s TR.

\vspace{-0.15in}
\subsection{Preprocessing}
The fMRI data were minimally preprocessed, including slice timing correction, despiking, registration, nonlinear transformation to MNI coordinate space, and masking of voxels outside of the brain. We used the Freesurfer parcellation to obtain 16 regions of interest (ROIs) associated with language and auditory processing: These ROIs include the central sulcus (CS), planum polare (PP), planum temporale (PT), postcentral gyrus (PoCG), precentral gyrus (PreCG), superior temporal gyrus (STG), transverse temporal gyrus (TTG) and transverse temporal sulcus (TTS) from both left and right hemispheres. These regions are commonly associated with auditory,  somatosensory and motor processing generally and speech perception more specifically. The mean BOLD time series were computed across voxels within each ROI.

\vspace{-0.1in}
\subsection{Shared Structure in Dynamic Connectivity Across Subjects}

To characterize dynamic FC, we computed time-varying correlations between the 16 ROI fMRI time series for each subject, using sliding windows of 15 TRs (22.5 s) with step size of 1 TR (1.5 s) between windows. The choice of window length of 22.5 s was based on the inverse of the lowest frequency present in the data \cite{Leonardi2015} after high-pass filtering at 0.05 Hz. This window length has been used in \cite{Bolton2018} to compute the sliding-window ISFC for movie-watching fMRI data.
At each time point $t$, the vectorized correlation coefficients of individual subjects are stacked as columns of matrix ${\bf Z}_t \in \mathbb{R}^{120 \times 13}$. We then performed the fused L+S decomposition on the multi-subject dynamic correlation matrices ${\bf Z}_1, \ldots, {\bf Z}_T$ by solving (\ref{Eq:FusedLS}) using the proposed ADMM algorithm. We selected the penalty parameter $\lambda_2$ based on the performance of the resulting low-rank models in predicting the time-varying auditory stimuli (see Section IV-E). From a range of $\lambda_2$ values from 0.001 to 0.5, we identified $\lambda_2 = 0.05$ as optimal, giving the highest classification accuracy on the validation set in a leave-one-subject-out cross-validation. Fig.~\ref{Fig:tv-analysis} shows the extracted time-varying low-rank and sparse components averaged across subjects and across all edges. In Fig.~\ref{Fig:tv-analysis} (left), the low-rank component shows noticeable time-varying structure with higher specificity in time compared to the original dynamic correlations. It provides a better localization of dynamic changes in connectivity which were possibly elicited by the movie and time-locked to the processing of specific ongoing movie events. From Fig.~\ref{Fig:tv-analysis} (right), it is apparent that the low-rank components are highly synchronized across subjects over time, while identifying some inter-subject variability possibly arising from varied responses of different individuals to the stimuli. In contrast, the sparse components show temporally uncorrelated patterns. The low-rank part is specifically sensitive to synchronized responses across subjects. This suggests that the proposed fused L+S decomposition is capable of isolating reliable FC changes driven by a correlated source across subjects (exposed to the same movie stimuli in our case). The uncorrelated background sources of dynamic FC across subjects are filtered out as the residual sparse components. These sources may correspond to subject-specific dynamic fluctuations of FC networks as well as non-neuronal artifacts unrelated to the stimulus-processing.

\begin{figure*}[!t]
	\begin{minipage}[t]{1\linewidth}
		\centering
		\includegraphics[width=1\linewidth,keepaspectratio]{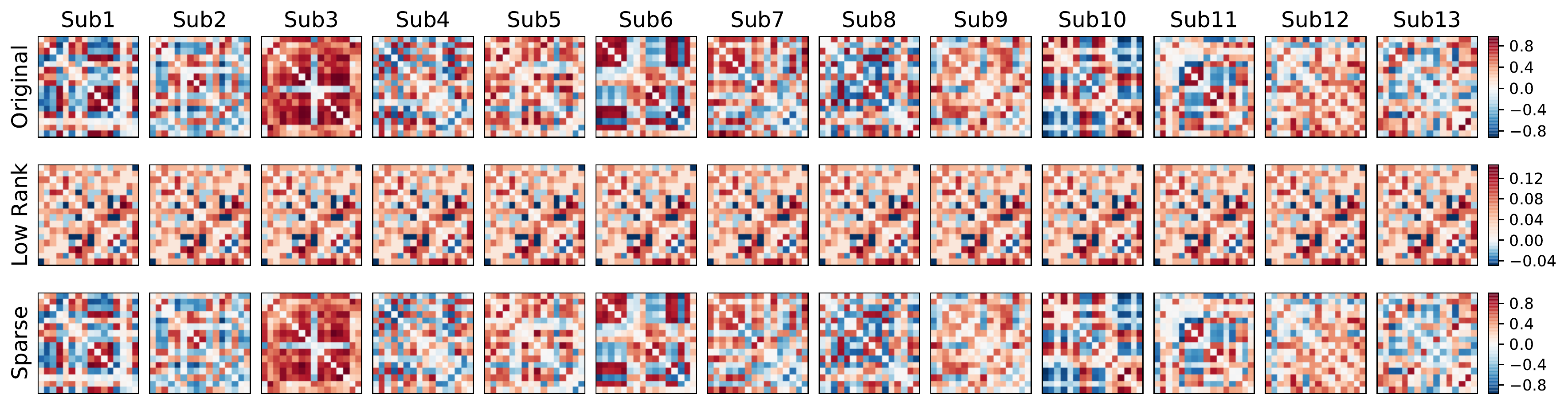}
	\end{minipage}
	\vspace{-0.25in}
\caption{Common structure in time-resolved functional connectivity across subjects. (Top) An example of original functional connectivity matrices ${\bf Z}_t =  [{\bf z}_{t1}, \ldots, {\bf z}_{tM}]$ for $M=13$ subjects at time window $t=59$ TRs. (Middle) Low-rank component $\widehat{\bf L}_t$ with $\hat{r}_t = 2$ recovered by our fused L+S approach, revealing shared connectivity structure across subjects. (Bottom) Estimated sparse errors or residuals $\widehat{\bf S}_t$ correspond to individual-specific background components. The selected time window corresponds to the median value of mean correlation over all edges and subjects.}
\label{Fig:LS-fmri-single-t-median}
	\vspace{-0.1in}
\end{figure*}

For illustration, Fig.~\ref{Fig:LS-fmri-single-t-median} shows decomposition for a snapshot of connectivity networks ${\bf Z}_t$ at time point $t=59$ TRs (selected based on median correlation over time). The low-rank component identifies a common network structure underlying all subjects, as evident from the similar and highly-correlated connectivity patterns detected across subjects. As validated by simulation, our algorithm can reliably recover the common structure from the subject-specific background noise, which could be arbitrarily large in magnitude and contaminated most of the network correlations (e.g., the dense residuals for subject 3). The estimated rank was $\hat{r}_t = 2$, where individual networks (columns of $\widehat{\bf L}$) can be expressed as a linear combination of $2$ common basis connectivity matrices with different mixing weights for each subject. See Supplementary Fig.~1 for additional example decomposition at $t=72$ TRs.

\vspace{-0.1in}
\subsection{Low-Rank Components Reveal Stimulus-Induced Connectivity Patterns}

\begin{figure}[!t]
\centering
\includegraphics[width=0.95\linewidth,keepaspectratio]{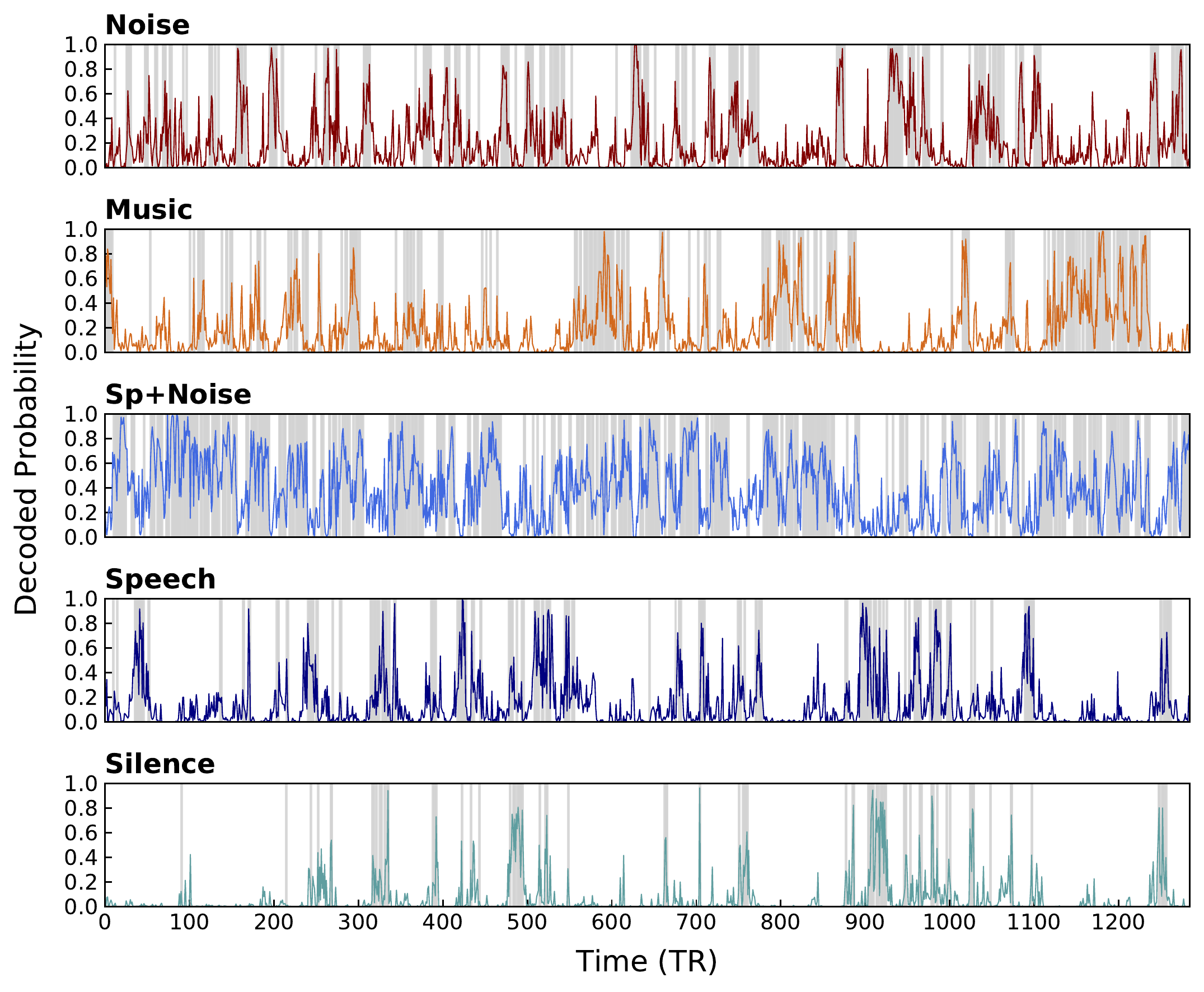}
\vspace{-0.1in}
	\caption{Decoded probability of a specific auditory stimulus category being present in the movie over time based on the multinomial logistic regression fitted on the extracted low-rank components. Colored lines indicate time courses of decoded probabilities averaged over all subjects. Gray regions show time points when the stimulus category was actually present in the movie.}  \label{Fig:decode-prob}
	\vspace{-0.1in}
\end{figure}

To test the aforementioned hypothesis, we examined whether the extracted low-rank components are related to the auditory stimuli present in the movie. We did so by fitting a multinomial logistic regression model to learn the mapping between the audio annotations and the time-resolved low-rank connectivity metrics. We set the auditory category labels as responses and the connectivity edges as predictors. The silence events were used as the reference category. To evaluate the goodness-of-fit and how well the low-rank components are temporally mapped to the auditory events, we used the fitted logistic model to decode the probability of the actual category labels at each time point. Fig.~\ref{Fig:decode-prob} shows the decoded time courses for each auditory category. For all categories, high probabilities are precisely aligned with the intervals when those categories are present, indicating accurate mapping of the low-rank components to the stimulus annotations.

\begin{figure*}[!t]
\centering
	\begin{minipage}[t]{0.85\linewidth}
		\centering
		\includegraphics[width=1\linewidth,keepaspectratio]{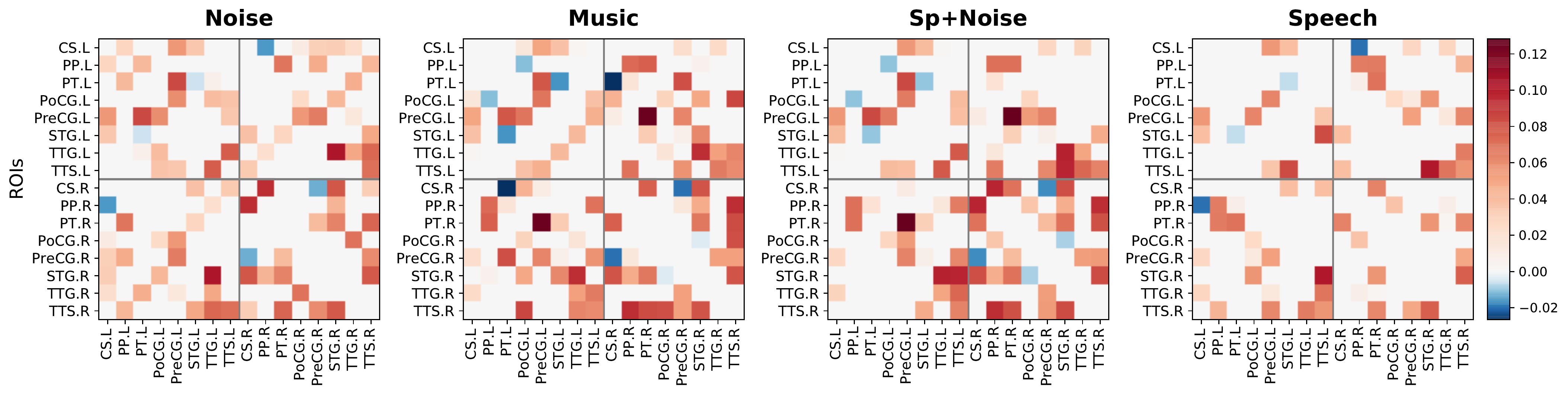}
	\end{minipage}
	\vspace{0.2 cm}
	\begin{minipage}[t]{0.85\linewidth}
		\centering
		\includegraphics[width=1\linewidth,keepaspectratio]{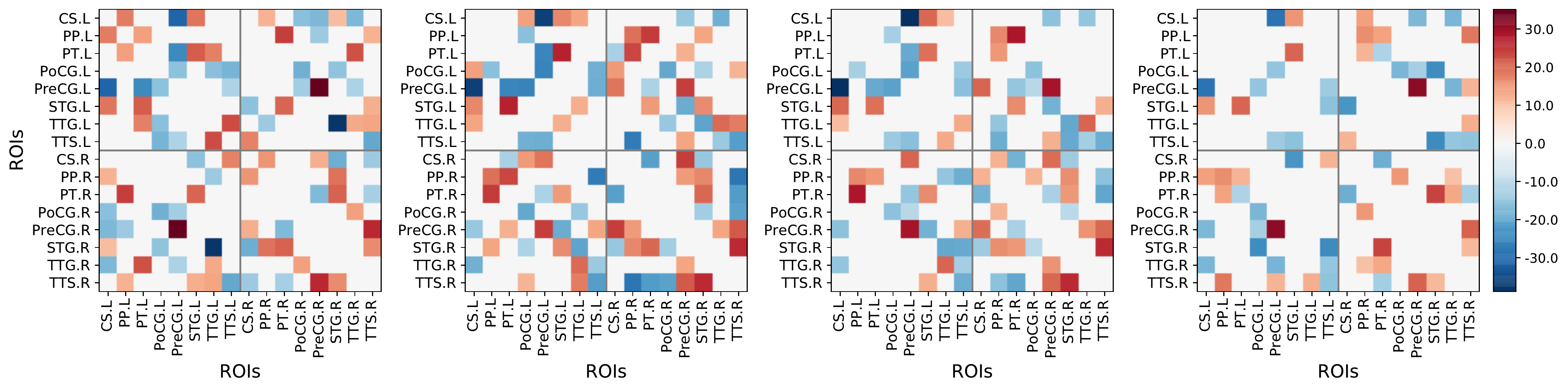}
	\end{minipage}
	\vspace{-0.2in}
\caption{Stimulus-induced connectivity networks during movie watching as revealed by the low-rank components. (Top) Average functional connectivity maps (across 13 subjects and time steps) associated with each of the four categories of auditory events in the natural movie. (Bottom) The corresponding spatial maps of regression coefficients from the fitted logistic model. Only functional connections showing significant stimulus-related effects are displayed (masked by regression coefficients with $p<0.0001$, Bonferroni-corrected for multiple testings).}
\label{Fig:mnr_net_lag-2}
\vspace{-0.1in}
\end{figure*}

\begin{figure}[!t]
\centering
	\begin{minipage}[t]{1\linewidth}
		\centering
		\subfloat[]{\includegraphics[width=1.02\linewidth,keepaspectratio]{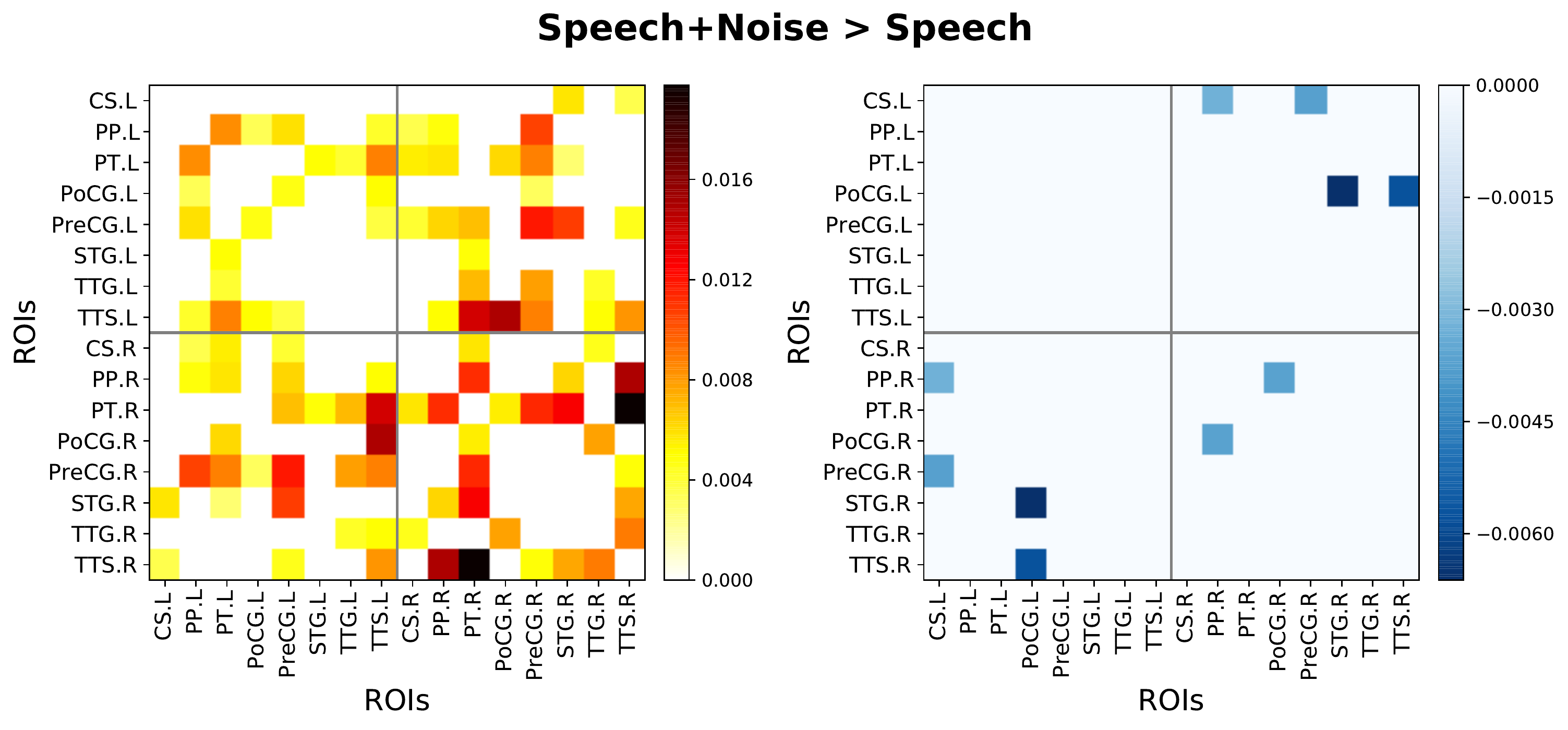}}
	\end{minipage}
	\begin{minipage}[t]{1\linewidth}
		\centering
		\subfloat[]{\includegraphics[width=0.75\linewidth,keepaspectratio]{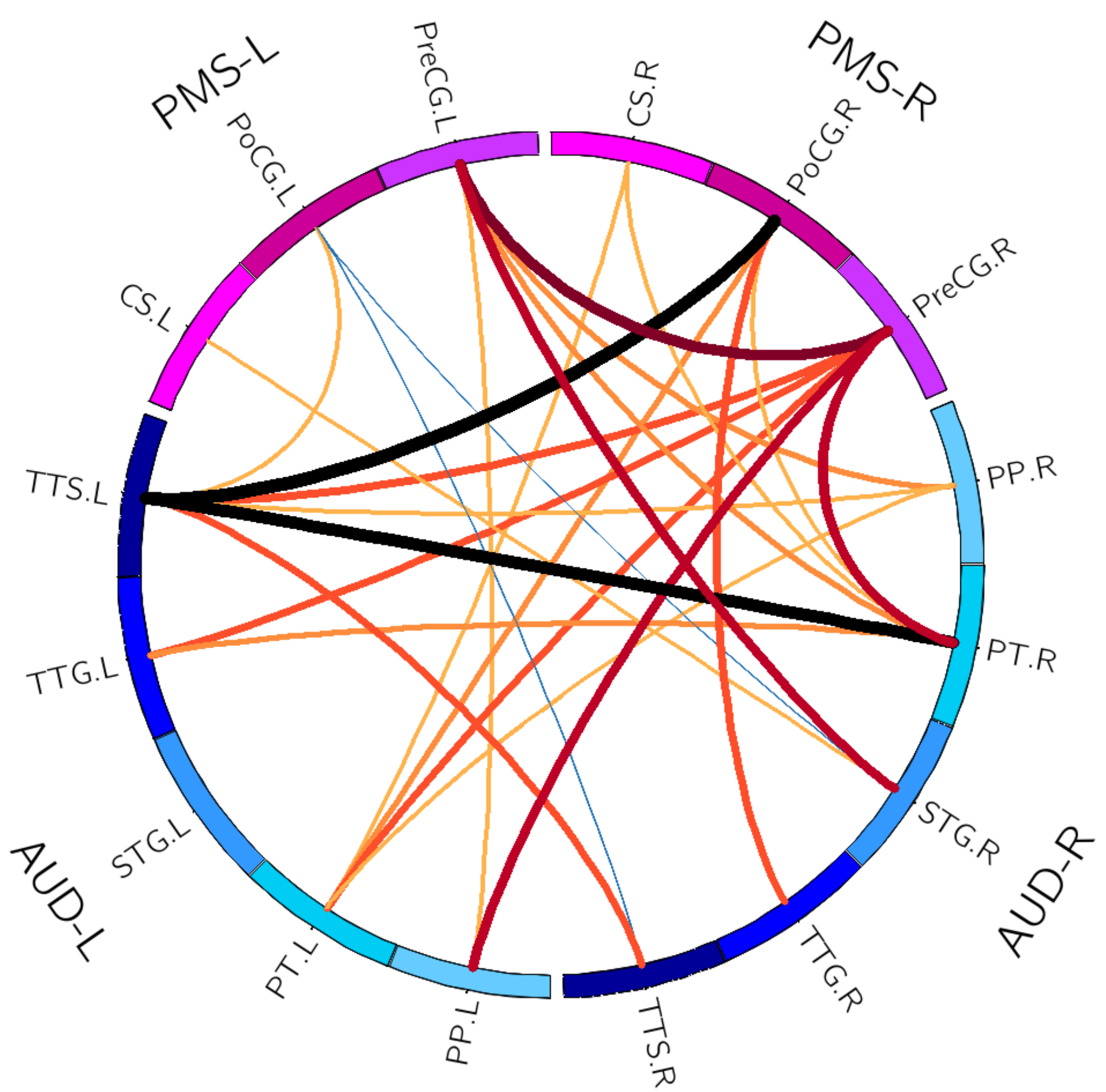}}
	\end{minipage}
	\vspace{-0.1in}
\caption{Difference in functional connectivity between the speech-in-noise (\textit{sp+noise}) and speech-only (\textit{speech}) events as revealed by the low-rank components at time lag of 3 s (or 2 TRs). (a) Connectivity maps show increase (left) and decrease (right) in FC during \textit{sp+noise} compared to \textit{speech} events. Only edges with significant difference in FC strength (absolute value of correlation coefficients) are shown ($p<0.0001$, Bonferroni-corrected). The five connections exhibiting largest increase in FC include PT.R-TTS.R, TTS.L-PoCG.R, PP.R-TTS.R, TTS.L-PT.R and PT.R-STG.R. (b) Connectogram shows changes increase (hot colors) and decrease (blue) in FC strength between auditory (AUD) and premotor-motor-somatosensory (PMS) regions. The thickness of links indicate magnitude of changes.}
\label{Fig:spn-vs-sp-lag2}
\end{figure}

In Fig.~\ref{Fig:mnr_net_lag-2}, we plot the mean functional connectivity maps and the estimated regression coefficient maps associated with the four auditory categories in the movie. The results are based on the low-rank components at a lag of 3 s (i.e., $\widehat{\bf L}_{t-2}$ at 2 time points before stimulus onsets at $t$). The latter was empirically determined to exhibit the strongest connectivity by testing at previous and future time points (at intervals of 1.5 s or 1 TR) relative to the stimulus onsets (See Supplementary Fig.~2 and Fig.~3). This lag was expected because it approximately includes to delay, rise time and peak of a canonical hemodynamic response function associated with any one stimulus lasting about 1.5 s. For details, see Supplementary Section III for the additional analyses on the temporal effect of brain response. As shown in Fig.~\ref{Fig:mnr_net_lag-2}, the low-rank component detected distinct connectivity patterns, with relatively different distributions of connectivity across the different auditory environments. As hypothesized, we observe a noticeable increase in connectivity when speech was accompanied by noise (\textit{sp+noise}), compared to speech in silence. We used a group-level t-test to contrast the FC between the \textit{sp+noise} and \textit{speech} conditions. Fig.~\ref{Fig:spn-vs-sp-lag2}(a) shows difference in the mean FC matrices (across subjects and time points) from the low-rank component between the two conditions. The connections were thresholded at $p<0.0001$ (Bonferroni-corrected). Specifically, there was significantly higher FC throughout regions typically involved in speech perception and language comprehension for speech-in-noise compared to speech. This included connections PT.R-TTS.R and PT.R-STG.R which showed the largest increase in FC strength. Furthermore, our method identified significant increase in connectivity between the regions of auditory and speech processing with the premotor, motor and somatosensory regions in response to speech mixed with noise, e.g., TTS.L-PoCG.R, PT.R-PreCG.R and TTG.R-PoCG.R, as shown in Fig.~\ref{Fig:spn-vs-sp-lag2}(b).

In summary, results support hypotheses about the role of `the motor system' in speech perception \cite{Adank2012,Skipper2017}. In particular, a distributed set of brain regions involved in producing speech are dynamically recruited in perceiving speech, particularly in more effortful listening situations. This might be because speech production regions are engaged in predicting upcoming speech sounds. In noisy environments, this process needs to be engaged more as predictions are inherently less accurate.

\vspace{-0.15in}
\subsection{Across-Subject Prediction of Stimulus Annotations}

We trained a support vector machine (SVM) classifier on the time-varying low-rank FC components to evaluate any improvement in predicting the different auditory categories over the original observed version of time-varying FC metrics. We also compared the performance with the inter-subject functional correlation (ISFC) \cite{Simony2016}, a widely-used method for isolating stimulus-induced inter-regional correlations between brains exposed to the same stimuli from other intrinsic and non-neuronal activities. We computed the ISFC in a sliding-window manner as in \cite{Bolton2018}, which showed that the time-varying ISFC can detect movie-driven fMRI FC changes by averaging out uncorrelated dynamic fluctuations. Following \cite{Simony2016}, we performed across-subject classification but using the time-resolved instead of static FC patterns to predict the stimulus events over time. We used leave-one-subject-out cross-validation. In each fold, one subject was held out for testing and the remaining $M-1$ subjects were used to train the SVM. We computed the time-resolved low-rank components ${\bf L}_t^{(m)} = [{\bf l}_{t,1}, {\bf l}_{t,m-1}, \ldots, {\bf l}_{t,m+1}, {\bf l}_{t,M}]$ for the training set from the standard FC, ${\bf Z}_t^{(m)} = [{\bf z}_{t,1}, {\bf z}_{t,m-1}, \ldots, {\bf z}_{t,m+1}, {\bf z}_{t,M}]$ using the fused L+S algorithm. Since the test set is unseen from the training set, we extracted the low-rank components for the test subject $m$ via orthogonal projection of the test data ${\bf l}_{t,m}={\bf U}{\bf U}^T{\bf z}_{t,m}$, where ${\bf U}$ is a $N^2 \times r$ basis matrix estimated based on the SVD of train data ${\bf L}_t^{(m)}$. Classification accuracy was computed as the percentage of time points correctly assigned to the true category labels.

\begin{figure}[!t]
	\begin{minipage}[t]{1\linewidth}
		\centering
		\includegraphics[width=0.75\linewidth,keepaspectratio]{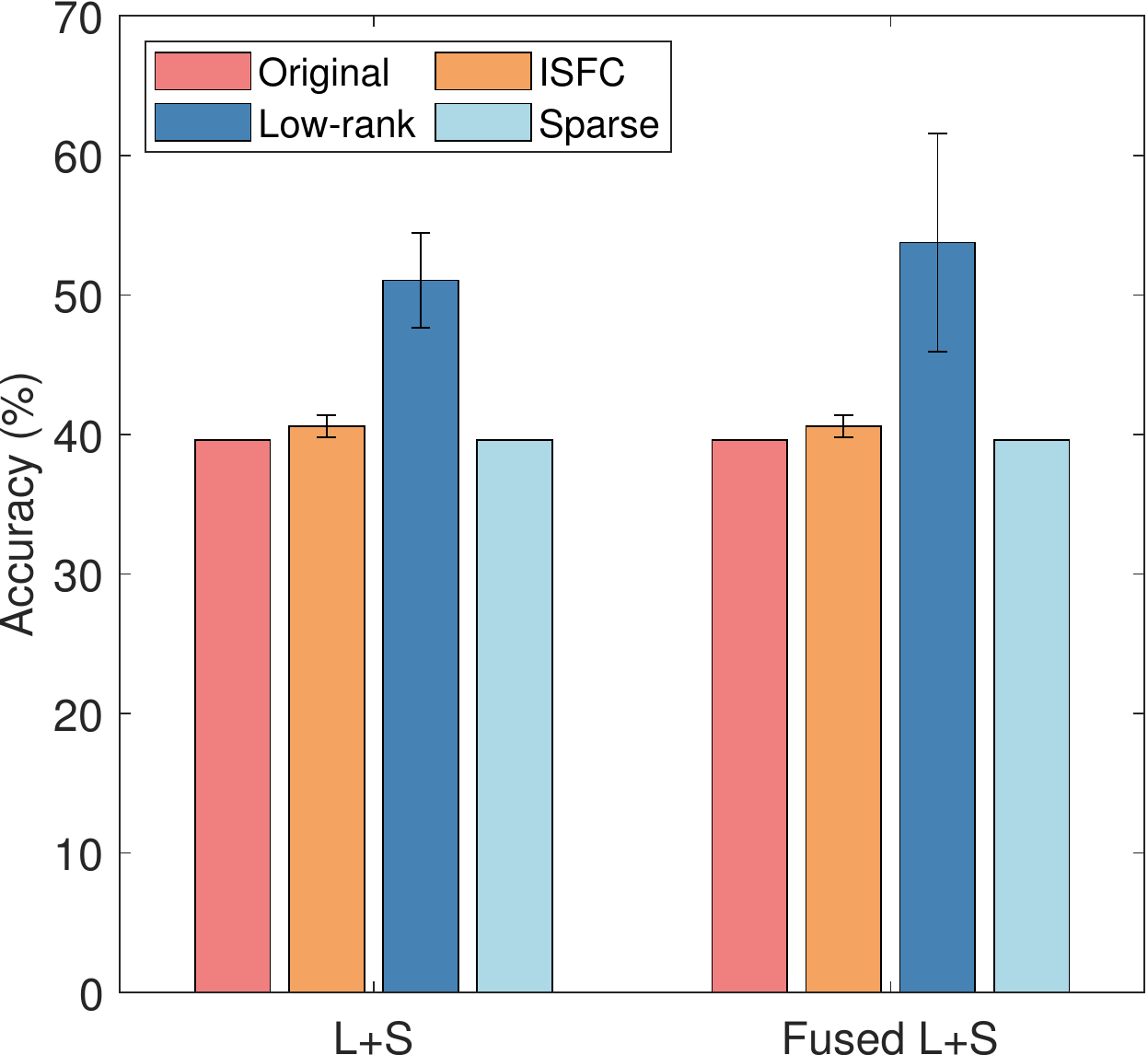}
	\end{minipage}
	\vspace{-0.25in}
\caption{Accuracy scores of leave-one-subject-out prediction of stimulus annotations for five categories of audio events present in the movie over time using different time-resolved connectivity features. The categories considered include \textit{Noise}, \textit{Music}, \textit{Speech}, \textit{Sp+Noise} and \textit{Silence}.}
\label{Fig:pred-acc}
	\vspace{-0.1in}
\end{figure}

Fig.~\ref{Fig:pred-acc} shows the classification results using the different time-varying FC metrics at a lag of 3 s relative to the stimulus annotations. Analysis of the temporal effects on the classification performance shows that the highest accuracy was obtained at this time lag (Supplementary Fig.~5). The low-rank components significantly outperformed both the standard FC and ISFC in discriminating ongoing auditory events in the movie. The ISFC is slightly better than the standard FC. The superior decoding performance suggests that the low-rank components are more stimulus-related, extracting novel information about the stimulus-induced dynamic functional changes not captured by the FC and ISFC. Among the L+S algorithms, addition of fused penalty improves the cross-subject prediction considerably, due to the further smoothing of subject-specific fluctuations to uncover shared neuronal responses that are time-locked to the same stimulus processing. As expected, the residual sparse components, corresponding to uncorrelated effects of intrinsic neuronal processes and noise, were poorly predictive of the stimuli across subjects.

\vspace{-0.05in}
\section{Conclusion}

We developed a novel L+S decomposition method for separating stimulus-induced and background components of dynamic FC networks during naturalistic neuroimaging, by leveraging on the time-locked nature of naturalistic stimuli. The method provides a principled way to recover both the stimulus-evoked, shared FC dynamics across individuals and the stimulus-unrelated idiosyncratic variations in individuals, respectively modeled as the low-rank and sparse structures of multi-subject dFC networks.
The proposed fused PCP solved via an efficient ADMM algorithm can capture the stimulus-induced similarities in FC profiles between subjects more effectively, and its robustness to dense noise allows us to filter out corruptions of FC edges that are not necessarily sparse. Our proposed framework is general, broadly applicable to other neuroimaging data such as electroencephalography (EEG), and can incorporate other FC measures beyond simple correlations in ISC-based analyses. In an application to movie-watching fMRI, our method identified time-locked changes in FC networks across subjects, which were meaningfully related to the processing of complex, time-varying auditory events in the movie, e.g., dynamic recruitment of speech production systems in perceiving speech in noisy environments. It also revealed potentially interesting individual differences in FC patterns that may relate to behavioral appraisal of stimuli. The extracted low-rank FC components also show better prediction of the movie auditory annotations than ISFC, suggesting the gain in information about the naturalistic stimuli by using shared network structure instead of synchronization of low-level regional activity. Thus, the proposed L+S recovery approach opens new opportunities for mapping brain network dynamics to stimulus features and behavior during naturalistic paradigms.
Future studies can explore potential applications of L+S analysis of natural fMRI to connectome-based prediction of individual traits and behavior, and diagnosis of cognitive and affective disorders. Further extensions are needed to improve the scalability of L+S learning algorithm to handle large-sized brain networks. The dynamic FC analysis can also be extended to investigate state-related changes in the network structure, e.g., dynamic community structure across subjects during naturalistic paradigms, by applying regime-switching models \cite{Ting2018,Ting2020} on the extracted time-resolved FC components.

\bibliography{Ref-LR+S}
\bibliographystyle{IEEEbib-3names}

\end{document}


\setlength{\abovedisplayskip}{7pt}
\setlength{\belowdisplayskip}{7pt}
\raggedbottom

\title{Supplementary Material}

\author{C.-M.~Ting, J.~I.~Skipper, S.~L.~Small, and~H.~Ombao}

\markboth{}%
{Shell \MakeLowercase{\textit{et al.}}: Bare Demo of IEEEtran.cls for Journals}

\maketitle

\vspace{-0.1in}
\begin{abstract}
This appendix contains additional material to accompany our paper ``Separating Stimulus-Induced and Background Components of Dynamic Functional Connectivity in Naturalistic fMRI". Section I provides some technical details of the proposed ADMM algorithm for fused L+S decomposition. Section II presents supplementary results for the detected shared and individual-specific connectivity structure from movie fMRI data. Section III investigates the temporal effect of brain response to auditory stimuli present in the movie.
\end{abstract}

\vspace{-0.05in}
\section{Technical Details: Update of ${\bf L}_t$}

We derive a closed-form solution for updating the low-rank component ${\bf L}_t$ in the proposed ADMM algorithm.

\begin{figure*}[!hb]
	\begin{minipage}[b]{1\linewidth}
		\centering
		\includegraphics[width=1\linewidth,keepaspectratio]{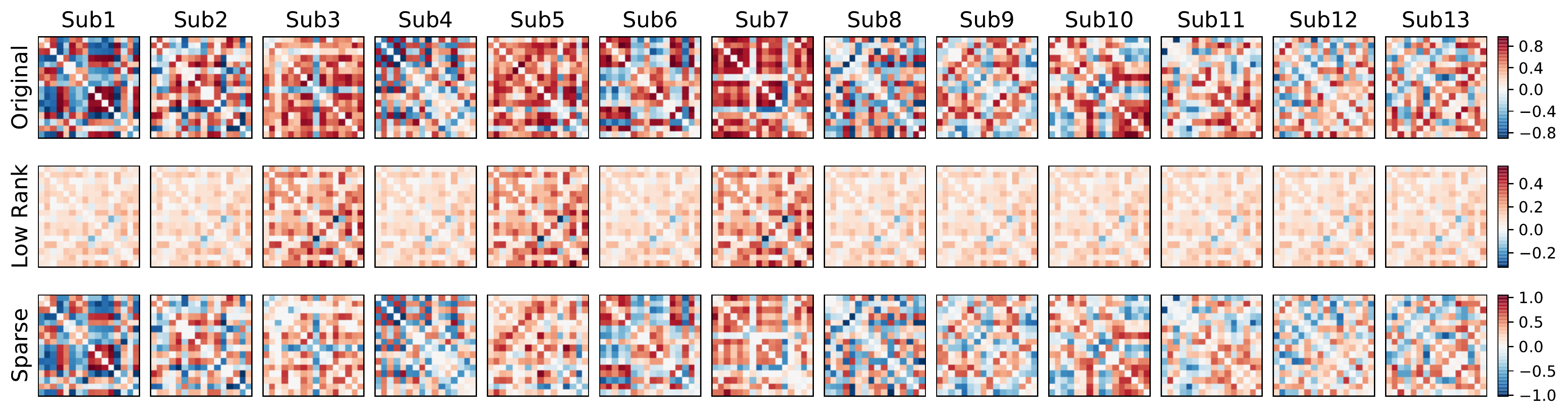}
	\end{minipage}
	\vspace{-0.5 cm}
\caption{Common structure in time-resolved functional connectivity across subjects. (Top) Example of original functional connectivity matrices ${\bf Z}_t =  [{\bf z}_{t1}, \ldots, {\bf z}_{tM}]$ for $M=13$ subjects at time window $t=72$ TRs. (Middle) Low-rank component $\widehat{\bf L}_t$ with $\hat{r}_t = 2$ recovered by our fused L+S approach, revealing shared connectivity structure across subjects. (Bottom) Estimated sparse errors or residuals $\widehat{\bf S}_t$ correspond to individual-specific background components. The time window selected corresponds to the maximum value of mean correlation over all edges and subjects.}
\label{Fig:LS-fmri-single-t-max}
	\vspace{-0.2in}
\end{figure*}

\begin{prop}
For any $\mu\geq0$, the solution (Eq.~16, main text) to the subproblem (15) as defined by the singular value thresholding (SVT) operator obeys
\begin{align}
&\mathcal{D}_{1/\mu(\nu+1)}({\bf Q}_t)  = \argmin_{{\bf L}_t} \left\{ \frac{1}{\mu} \left\|{\bf L}_t\right\|_{*} \right. \notag \\ & \left. + \frac{1}{2} \left\|{\bf L}_t+{\bf S}_t-{\bf Z}_t+{\bf X}_t \right\|_{F}^2  + \frac{\nu}{2} \left\|{\bf L}_t - {\bf C}_t \right\|_{F}^2 \right\} \label{Eq:prop1}
\end{align}
where ${\bf Q}_t = \frac{1}{\nu+1}({\bf Z}_t - {\bf S}_t + \nu{\bf C}_t - {\bf X}_t)$.
\end{prop}
\begin{proof}
We generally follow the lines of proof of Theorem 2.1 in \cite{Cai2010}. Let $\phi = \nu+1$ and $\tau = \mu^{-1}$. By the optimality condition of the objective function in (\ref{Eq:prop1}), the solution ${\bf L}_t^{*}$ must satisfy
\begin{equation}
{\bf 0} \in \phi{\bf L}_t^{*} - \boldsymbol{\Gamma}_t + \tau \partial\left\|{\bf L}_t^{*}\right\|_{*} \label{Eq:kkt}
\end{equation}
where $\partial\left\|{\bf L}_t^{*}\right\|_{*}$ is the subgradient of nuclear norm evaluated at ${\bf L}_t^{*}$, and $\boldsymbol{\Gamma}_t = {\bf Z}_t - {\bf S}_t + \nu{\bf C}_t - {\bf X}_t$. Let ${\bf M} = {\bf U}\boldsymbol{\Sigma}{\bf V}^T$ be the SVD of an arbitrary matrix ${\bf M}$. It is known that
\begin{equation}
\partial\left\|{\bf M}\right\|_{*} = \left\{{\bf U}{\bf V}^T + {\bf W}:  {\bf U}^T{\bf W}=0, \ {\bf W}{\bf V}=0, \ \left\|{\bf W}\right\|_2\leq1 \right\}. \notag
\end{equation}

Define ${\bf L}_t^{*} = \mathcal{D}_{\tau/\phi}(\phi^{-1}\boldsymbol{\Gamma}_t)$. To show that ${\bf L}_t^{*}$ obeys (\ref{Eq:kkt}), decompose the SVD of $\boldsymbol{\Gamma}_t$ as
\begin{equation}
\boldsymbol{\Gamma}_t = {\bf U}_0\boldsymbol{\Sigma}_0{\bf V}_0^T + {\bf U}_1\boldsymbol{\Sigma}_1{\bf V}_1^T \notag
\end{equation}
where ${\bf U}_0,{\bf V}_0$ (resp. ${\bf U}_1,{\bf V}_1$) are singular vectors with singular values $\sigma_{0,j} > \tau$ (resp. $\sigma_{1,j} \leq \tau$). It follows that the SVD of $\phi^{-1}\boldsymbol{\Gamma}_t$ is
\begin{equation}
\phi^{-1}\boldsymbol{\Gamma}_t = {\bf U}_0\widetilde{\boldsymbol\Sigma}_0{\bf V}_0^T + {\bf U}_1\widetilde{\boldsymbol\Sigma}_1{\bf V}_1^T \notag
\end{equation}
where $\widetilde{\boldsymbol\Sigma}_0 = \phi^{-1}\boldsymbol{\Sigma}_0$ (resp. $\widetilde{\boldsymbol\Sigma}_1 = \phi^{-1}\boldsymbol{\Sigma}_1$) with singular values $\widetilde\sigma_{0,j} > \tau/\phi$ (resp. $\widetilde\sigma_{1,j} \leq \tau/\phi$).
Then, we have
\begin{equation}
{\bf L}_t^{*} = {\bf U}_0 (\widetilde{\boldsymbol\Sigma}_0 - \tau/\phi{\bf I}){\bf V}_0 = \phi^{-1}{\bf U}_0 \left(\boldsymbol{\Sigma}_0 - \tau{\bf I} \right){\bf V}_0. \notag
\end{equation}
Therefore,
\begin{equation}
\boldsymbol{\Gamma}_t - \phi{\bf L}_t^{*} = \tau\left({\bf U}_0{\bf V}_0^T + {\bf W}\right), \ \ {\bf W}=(\tau/\phi)^{-1}{\bf U}_1\widetilde{\boldsymbol\Sigma}_1{\bf V}_1^T. \notag
\end{equation}
By definition, ${\bf U}_0^T{\bf W}=0$, ${\bf W}{\bf V}_0=0$ and since the diagonal elements of $\widetilde{\boldsymbol\Sigma}_1$ have magnitude bounded by $\tau/\phi$ and $\left\|{\bf W}\right\|_2\leq1$. Hence $\boldsymbol{\Gamma}_t - \phi{\bf L}_t^{*} = \tau \partial\left\|{\bf L}_t^{*}\right\|_{*}$.
\end{proof}

\vspace{-0.05in}
\section{Shared \& Subject-Specific FC Structure}

Fig.~\ref{Fig:LS-fmri-single-t-max} shows L+S decomposition for a snapshot of FC networks at time window $t=72$ (corresponding to the maximum value of correlation over time). The low-rank components detected shared FC patterns across subjects while revealing stronger response to the stimulus in subjects 3, 5 and 7.

\vspace{-0.05in}
\section{Temporal Effect of Brain Response}

\begin{figure*}[!t]
\centering
\includegraphics[width=1\linewidth,keepaspectratio]{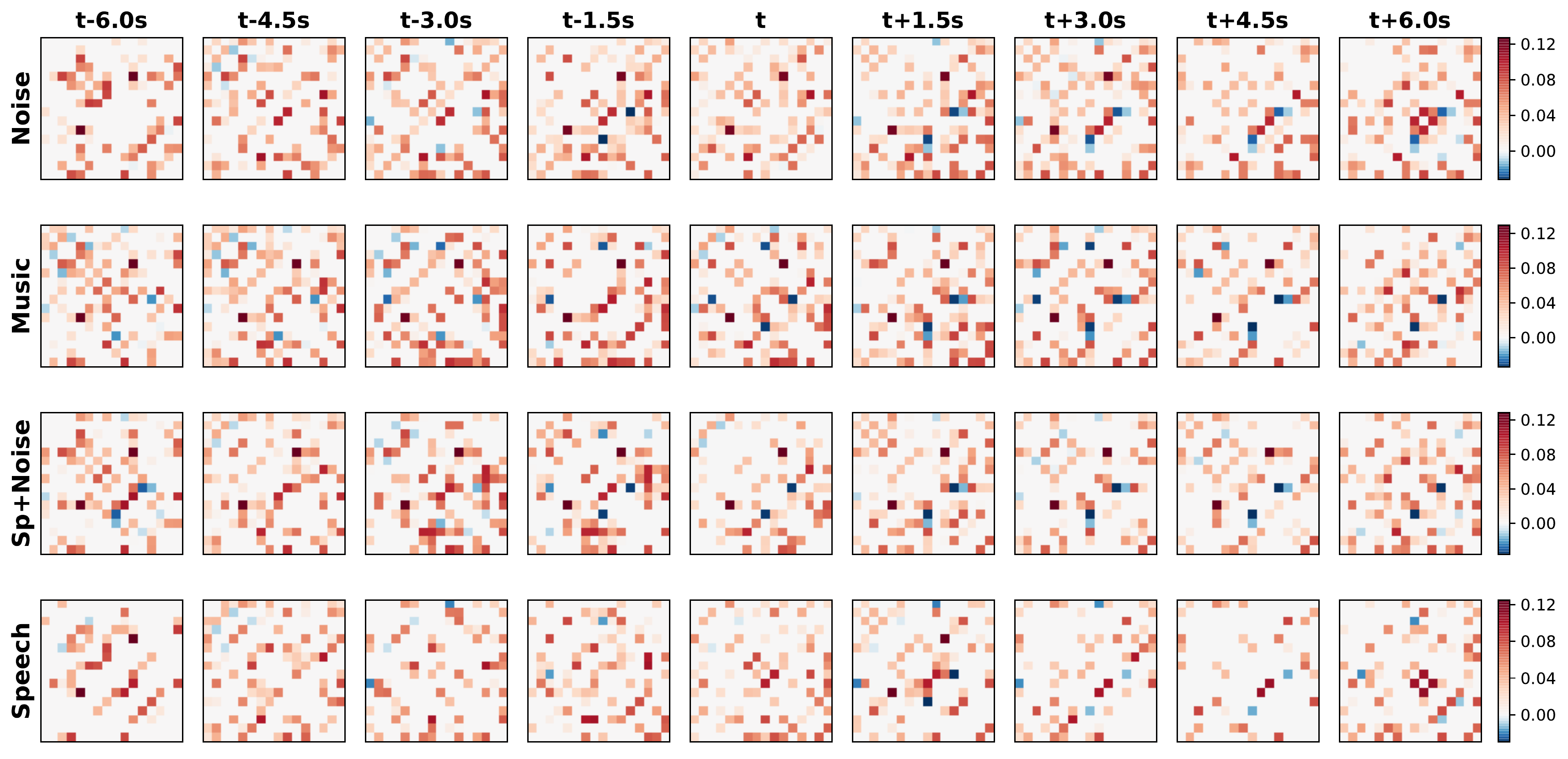} 
\vspace{-0.25in}
	\caption{Mean functional connectivity maps of the low-rank components (across subjects and time) at different time points before and after the stimulus onsets for four auditory events in the movie. The maps are masked by significant regression coefficients in the logistic model ($p<0.0001$, Bonferroni-corrected).}  \label{Fig:mnr-net-fc-lag}
	\vspace{-0.1in}
\end{figure*}

\begin{figure*}[!t]
\centering
\includegraphics[width=1\linewidth,keepaspectratio]{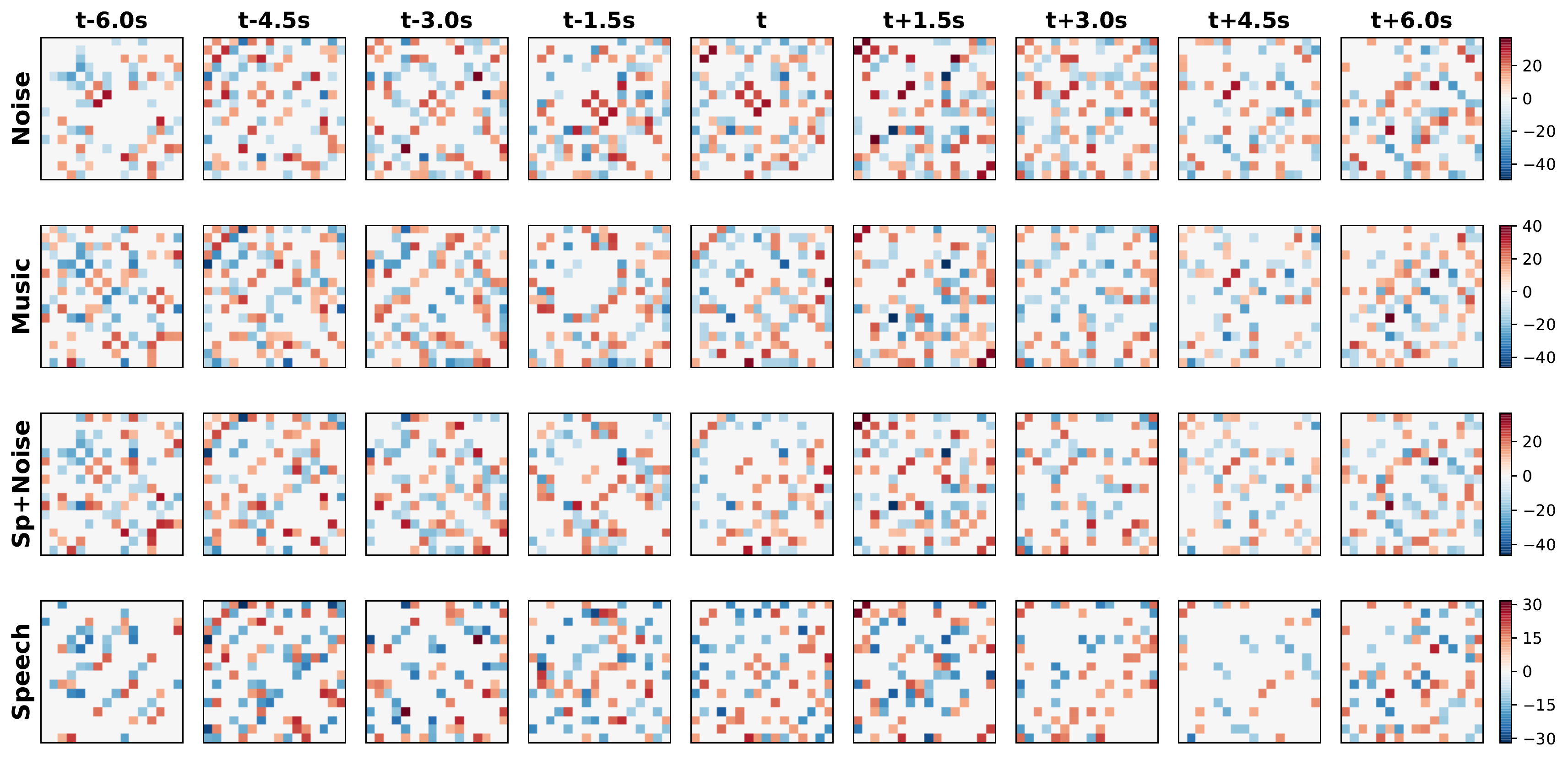} 
\vspace{-0.25in}
	\caption{Regression coefficient maps of logistic regression models fitted on time-resolved low-rank components at different time points before and after the stimulus annotations for four auditory events in the movie. The maps are thresholed at $p<0.0001$ (Bonferroni-corrected).}  \label{Fig:mnr-net-beta-lag}
	\vspace{-0.1in}
\end{figure*}

\begin{figure*}[!t]
 \centering
 \includegraphics[width=1\linewidth,keepaspectratio]{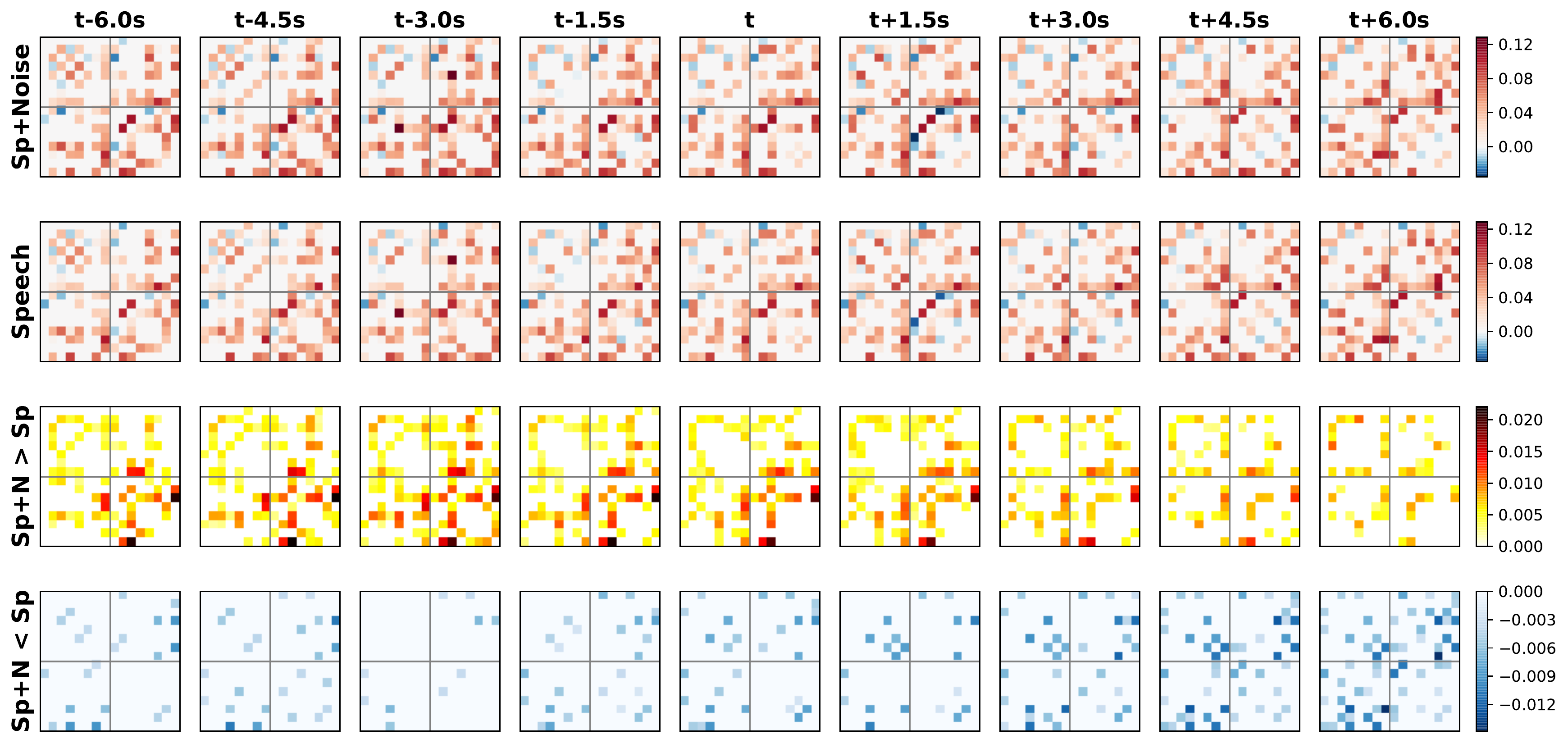}
 \vspace{-0.25in}
	 \caption{Contrast of functional connectivity between the speech-in-noise and speech events as revealed by the low-rank components at different time points before and after the stimulus onsets. Mean FC maps across subjects and time for the speech-in-noise and speech events (top panels) and their differences (bottom panels). Only edges with significant difference in FC strength (absolute value of correlation coefficients) are shown ($p<0.0001$, Bonferroni-corrected).}  \label{Fig:spn-vs-sp-hot}
\end{figure*}

To investigate the timing of brain response to stimuli, we examined the estimated functional connectivity (FC) before and after the onsets of auditory stimuli. Fig.~\ref{Fig:mnr-net-fc-lag} and Fig.~\ref{Fig:mnr-net-beta-lag} show respectively the FC and regression coefficient maps fitted based on low-rank components at future and previous time points (at intervals of 1.5 s or 1 TR) relative to the stimulus onsets. The responsiveness of brain networks varies as a function of temporal context. We observed that the strongest connectivity occurred 3 s prior to the auditory events (t-3 s), which was especially evident for the \textit{sp+noise} and \textit{music} categories. In Fig.~\ref{Fig:spn-vs-sp-hot}, we also contrast the FC for \textit{sp+noise} vs \textit{speech} categories at the different time points around stimulus onsets. Similarly, the largest increase in connectivity strength was found at the lag of 3 s before the onsets when perceiving speech-in-noise, with enhanced interactions between regions from the sensory motor cortex and the superior temporal plane. These interactions decreased gradually afterward over time with little activity after the stimulus onsets.

\begin{figure*}[!t]
	\begin{minipage}[t]{1\linewidth}
		\centering
		\includegraphics[width=0.5\linewidth,keepaspectratio]{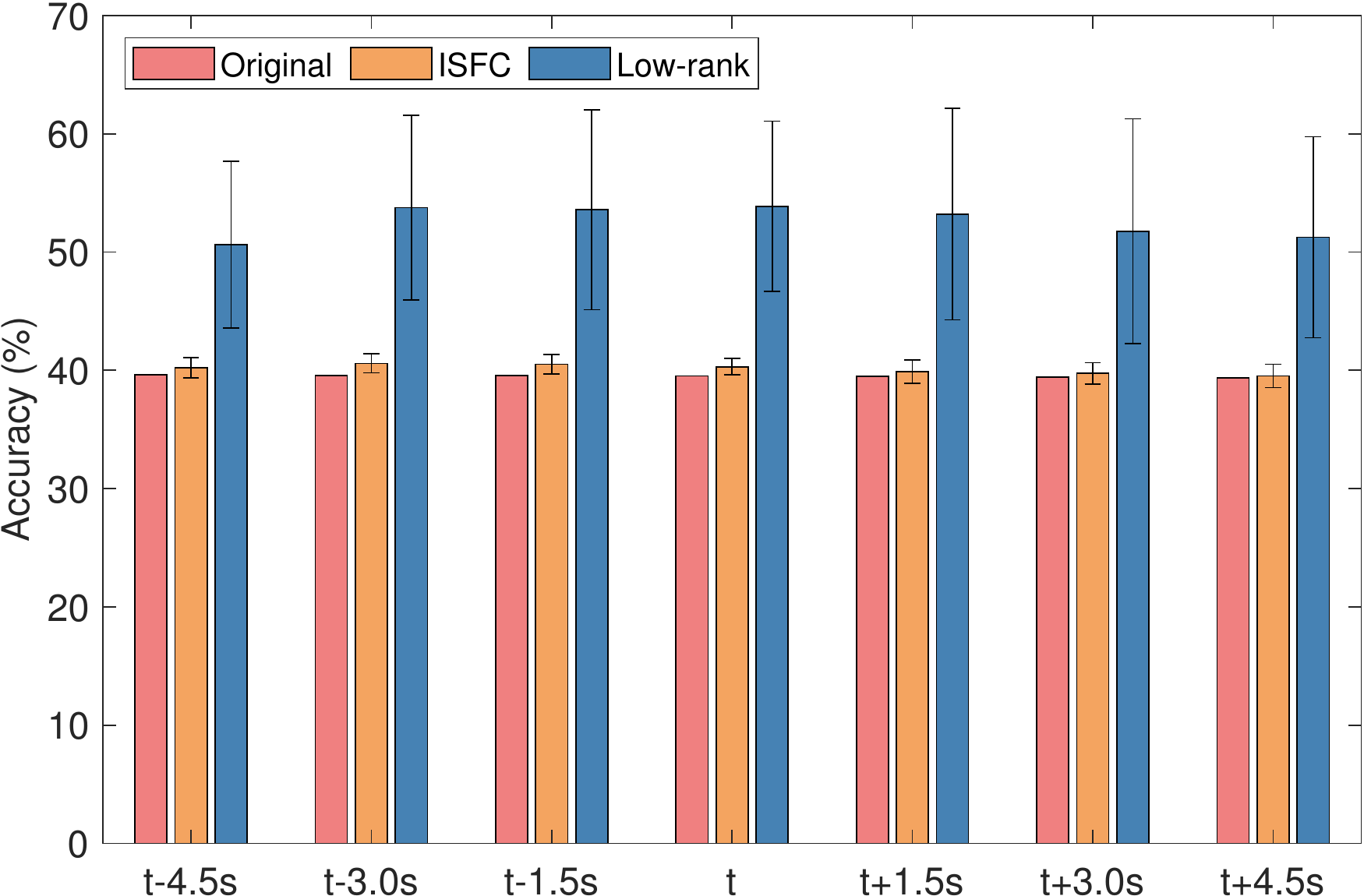}
	\end{minipage}
	\vspace{-0.2in}
\caption{Prediction accuracies of various time-resolved connectivity features at different time points before and after the auditory stimulus onsets. Error bars indicate standard deviations of accuracy across subjects in leave-one-subject-out cross-validation.}
\label{Fig:time-pred-acc}
\end{figure*}

To examine the temporal effect on the cross-subject prediction of stimulus annotations, we trained the SVM classifiers using the preceding and succeeding FC patterns to predict the auditory events. Fig.~\ref{Fig:time-pred-acc} shows that use of the proposed fused low-rank FC performed consistently better than the standard FC and ISFC at all time points considered before and after the stimulus onsets. The highest prediction accuracy was achieved at 3 s before the onsets suggesting the FC metrics at this time lag provided the most discriminative information about the stimuli. This is confirmed by the largest contrast in FC pattern between stimulus types detected at this time lag (Fig.~\ref{Fig:spn-vs-sp-hot}). The drop in prediction performance when using the succeeding FC implies deceased level of brain responsiveness to the stimuli over time after the onsets.

\bibliography{Ref-LR+S}
\bibliographystyle{IEEEbib-3names}